\documentclass[amsmath,amssymb,amsbsy,reprint,prb,preprintnumbers,showpacs,superscriptaddress,longbibliography]{revtex4-2}
\usepackage{graphicx,color}
\usepackage{dcolumn}
\usepackage{bm}
\usepackage{braket}
\usepackage{mathtools}
\usepackage[normalem]{ulem}
\usepackage[breaklinks,colorlinks=true,linkcolor=blue,urlcolor=blue,citecolor=blue]{hyperref}
\usepackage{times}
\usepackage{physics}
\usepackage{latexsym}
\usepackage{amsmath, amssymb}
\usepackage{mathtools}
\usepackage{multirow}

\newcommand{\be}{\begin{eqnarray}}
\newcommand{\ee}{\end{eqnarray}}

\renewcommand{\theequation}{\arabic{equation}}

\begin{document}

\title{Entanglement Asymmetry and Asymmetry Contour under Boundary Symmetry Breaking}
\date{\today}
\author{Takahiro Orito}
\affiliation{Department of Physics, College of Humanities and Sciences, Nihon University, Sakurajosui, Setagaya, Tokyo 156-8550, Japan}
\author{Daisuke Yamamoto}
\affiliation{Department of Physics, College of Humanities and Sciences, Nihon University, Sakurajosui, Setagaya, Tokyo 156-8550, Japan}
\affiliation{Quantum Many-Body Dynamics Research Team, RIKEN Center
for Quantum Computing (RQC), Wako, Saitama 351-0198, Japan
}
\affiliation{Global R\&D Center for Business by Quantum-AI Technology (G-QuAT),
National Institute of Advanced Industrial Science and Technology (AIST),
1-1-1 Umezono, Tsukuba, Ibaraki 305-8568, Japan
}

\begin{abstract}
We investigate how symmetry-breaking boundary magnetic fields affect magnetization and R\'enyi-2 entanglement asymmetry (EA) in the transverse-field Ising model.
For the ground state, magnetization exhibits qualitatively distinct behaviors in the ordered and disordered phases.
EA also exhibits phase-dependent behavior, which is well captured by the lower and upper bounds derived in this work.
In the critical regime, both magnetization and EA exhibit behavior qualitatively consistent with the field-theoretical predictions for a related setup.
We also study boundary-quench dynamics and find that magnetization propagates from the boundary into the bulk with a velocity consistent with the maximum quasiparticle velocity.
In contrast, because EA is a subsystem-integrated quantity, its spatial structure is not directly resolved.
We therefore introduce the asymmetry contour, a contour-like finite-difference quantity that provides a spatially resolved characterization of the symmetry-breaking information encoded in the reduced density matrix. The resulting contour is complementary to the local magnetization, retaining sensitivity to nonlocal many-body correlations while its dynamics also reflects the underlying quasiparticle propagation. 
These results show how EA and its contour can provide complementary information-theoretic characterizations of symmetry breaking and its spatial structure.
\end{abstract}


\maketitle
\section{Introduction} 
Symmetry and its breaking play pivotal roles in quantum many-body physics~\cite{wen2004quantum,sachdev2011quantum}.
A paradigmatic manifestation of symmetry breaking arises when a Hamiltonian is invariant under a given symmetry, whereas its equilibrium phase can exhibit a lower symmetry.
This mismatch is conventionally understood within the framework of spontaneous symmetry breaking, and the resulting symmetry-breaking pattern is characterized by an appropriate order parameter.
Recent developments in nonequilibrium physics~\cite{Sieberer_2025,Harper_2020}, open quantum systems~\cite{Fazio_2025,Buca_2012}, and quantum information~\cite{de_Groot_2022} have further highlighted the significance of symmetry and its breaking beyond equilibrium phase classification~\cite{Kawabata_2019,Xiao_2025,wang_2026}. 

Beyond identifying which symmetry is present and whether it is broken, it is also important to quantify the degree of symmetry breaking.
Entanglement asymmetry (EA) has recently emerged as a useful information-theoretic measure for this purpose~\cite{Ares2023}.
Rather than characterizing symmetry breaking of the entire system, the EA quantifies the extent of symmetry breaking  within a specific subsystem.
Since its introduction, the nonequilibrium behavior of EA has been extensively investigated in the global quench protocols, where an initially symmetry-breaking state evolves under a symmetry-preserving Hamiltonian~\cite{Yamashika2024,Rylands2024,Murciano2024,Ares2025,Yamashika2025,Hara2026}.
The primary focus of these studies is dynamical symmetry restoration and the associated quantum Mpemba effect: the more strongly the symmetry is broken, the faster it is restored~\cite{Ares2023}.

Despite this remarkable progress, most previous studies have focused on restoring symmetry.
In contrast, how locally induced symmetry breaking develops and propagates through a system remains largely unexplored~\cite{Yu2025,fossati2026}.
More recently, boundary-induced symmetry breaking has begun to be explored using boundary conformal field theory (BCFT)~\cite{fossati2026}.
However, it remains unclear how boundary-induced symmetry breaking extends into the bulk and how its behavior varies depending on the underlying quantum phases.

In this study, we address these questions using a tractable lattice setup in which a symmetry-breaking perturbation at the boundary can be continuously varied and EA can be efficiently evaluated.
Taking the transverse-field Ising model as an example, we achieve this by introducing ancilla spins, which allow us to evaluate R\'enyi-2 EA efficiently.
Using this formulation, we investigate both the ground-state properties and the nonequilibrium dynamics induced by the sudden introduction of symmetry-breaking boundary fields.

In the ground state, the EA exhibits distinct phase-dependent behavior consistent with the bounds derived in this work, while its critical behavior is qualitatively consistent with field-theoretical predictions for a related setup~\cite{Fossati2025}.
We then turn to the nonequilibrium dynamics, comparing the EA with the local magnetization to clarify the different aspects of symmetry breaking captured by these quantities.
Because EA is a subsystem-integrated quantity, once a subsystem contains a region carrying considerable symmetry-breaking information, it becomes less sensitive to the subsequent spatial spread of that information.
This motivates us to introduce a contour-like finite-difference quantity, the asymmetry contour (AC), to provide a spatially resolved characterization of the symmetry-breaking information captured by entanglement asymmetry.
We find that the dynamics of AC reflects the underlying quasiparticle propagation while retaining sensitivity to the broader symmetry-breaking information contained in the reduced density matrix, beyond that captured by the local magnetization alone.

The rest of this paper is organized as follows. 
In Sec.~II, we introduce the transverse-field Ising chain with symmetry-breaking boundary magnetic fields at both ends.
We then formulate the problem in terms of free fermions by introducing two ancilla spins.
In Sec. III, we investigate the ground-state properties.
In Sec. IV, we investigate the dynamics following a sudden quench of boundary magnetic fields. 
Section V is devoted to the conclusion.

\section{The quantum Ising chain with symmetry-breaking boundary magnetic fields}
We consider the quantum Ising chain with symmetry-breaking boundary magnetic fields:
\begin{eqnarray}
\hat{H}=\hat{H}_0-h_bX_{1}-h_bX_{L},
\label{TFIM_B}
\end{eqnarray}
where 
\begin{eqnarray}
\hat{H}_0=-\sum^{L-1}_{i=1}JX_{i}X_{i+1}-\sum^{L}_{i=1}hZ_{i}
\label{TFIM}
\end{eqnarray}
represents the one-dimensional transverse-field Ising model (TFIM) with open boundary conditions~\cite{Mbeng2024}.
Here, $Z_i$ and $X_i$ are Pauli operators at site $i$ while $J$ denotes the nearest-neighbor coupling, $h$ is the bulk transverse field, and $h_b$ is the symmetry-breaking boundary field.
Note that Ref.~\cite{Fossati2025} focuses on the limit $h_b\to\infty$, whereas we consider finite $h_b$. 

Although the Hamiltonian~(\ref{TFIM}) can be mapped onto a quadratic fermionic Hamiltonian, the boundary magnetic fields in model~(\ref{TFIM_B}) prevent the full Hamiltonian from being transformed into a quadratic form, making a direct analytical treatment difficult. 
Nevertheless, it has already been shown that the problem can be recast as a quadratic fermionic problem~\cite{Campostrini2015} by introducing two additional ancilla spins, one at each end of the chain. 
The Hamiltonian in the enlarged Hilbert space is given by
\begin{eqnarray}
\hat{H}_E=\hat{H}_0-h_bX_{0}X_{1}-h_bX_{L}X_{L+1}.
\label{TFIM_E}
\end{eqnarray}

Here, we explain the relation between the enlarged Hamiltonian~(\ref{TFIM_E}) and the original Hamiltonian~(\ref{TFIM_B}).
Let us first clarify the role of the ancilla spins $X_{0}$ and $X_{L+1}$.
Since $X_{0}$ and $X_{L+1}$ commute with each other and with the Hamiltonian~(\ref{TFIM_E}), they can be diagonalized simultaneously. 
Then, by using the eigenvalues of $X_{0}$ and $X_{L+1}$, denoted by $(\delta_0,\delta_{L+1})$, the enlarged Hilbert space can be divided into four sectors: $(\delta_0=\pm 1,\delta_{L+1}=\pm 1)$.
It is now clear that $\delta_0$ and $\delta_{L+1}$ act as sign variables for the boundary magnetic fields:
\begin{eqnarray}
\hat{H}_E=\hat{H}_0-h_b\delta_{0}X_{1}-h_b\delta_{L+1}X_{L}.
\end{eqnarray}
Therefore, selecting the appropriate $\delta_0$ and $\delta_{L+1}$ is essential to relate the enlarged Hamiltonian~(\ref{TFIM_E}) to the original Hamiltonian~(\ref{TFIM_B}).

\subsection{Implementation of the ancilla-spin approach}
We will explain how to implement the ancilla-spin approach introduced above using a fermionic representation. 
First, we concentrate on the quadratic structure of the Hamiltonian $\hat{H}_E$.
We then discuss how to select the appropriate sectors of $X_0$ and $X_{L+1}$. 

\subsubsection{Jordan-Wigner transformation and quadratic Hamiltonian}
Using the Jordan-Wigner transformation,  
\begin{eqnarray}
X_i=K_i(c_i^\dagger+c_i),\ Y_i=iK_i(c_i^\dagger-c_i),\ {\rm and}\ Z_i=(1-2n_i),\nonumber\\
\label{JWT}
\end{eqnarray}
where $K_i=\prod_{j=0}^{i-1}e^{i\pi n_j}$, $n_j=c_j^\dagger c_j$, and $c_i$ ($c_i^\dagger$) is a fermionic annihilation (creation) operator, the enlarged Hamiltonian is mapped onto a quadratic fermionic Hamiltonian:
\begin{eqnarray}
\hat{H}_{{\rm EF}}&=&\hat{H}_0+\hat{H}_B
\label{TFIM_EF}
\end{eqnarray}
where
\begin{eqnarray}
\hat{H}_{0}&=&-\frac{J}{2}\sum_{i=1}^{L-1}[c_i^\dagger c_{i+1}+c_i^\dagger c_{i+1}^\dagger+h.c.]+h\sum_{i=1}^{L}[c_i^\dagger c_{i}-c_{i}c_i^\dagger].\nonumber\\
\label{TFIM0_EF}
\end{eqnarray}
and
\begin{eqnarray}
\hat{H}_{B}&=&
-\frac{h_b}{2}[c_0^\dagger c_{1}+c_0^\dagger c_{1}^\dagger+h.c.]
-\frac{h_b}{2}[c_L^\dagger c_{L+1}+c_{L}^\dagger c_{L+1}^\dagger+h.c.].\nonumber\\
\label{TFIMB_EF}
\end{eqnarray}
The Hamiltonian~(\ref{TFIM_EF}) can be written in the Bogoliubov–de Gennes (BdG) form with a $(2L+4)\times(2L+4)$ matrix as
\begin{eqnarray}
H_{\rm EF}&=&
\begin{pmatrix}
   c^\dagger & c 
\end{pmatrix}
H_{{\rm BdG}}
\begin{pmatrix}
   c  \\
   c^\dagger 
\end{pmatrix},
\label{TFIM_EF_MAT}
\end{eqnarray}
where $\begin{pmatrix}
   c^\dagger & c 
\end{pmatrix}=\begin{pmatrix}
   c_0^\dagger &c_1^\dagger &\cdots & c_{L+1}^\dagger & c_0 &c_1 &\cdots & c_{L+1} 
\end{pmatrix},$
and
\begin{eqnarray}
H_{{\rm BdG}}&=&
\begin{pmatrix}
   A & B\\
   -B & -A
\end{pmatrix}.
\label{TFIM_BDG_MAT}
\end{eqnarray}
Here, the matrices A and B are given by
\begin{eqnarray}
A&=&
\begin{pmatrix}
   0 & -\frac{h_b}{2}&0&\cdots& \cdots&0\\
   -\frac{h_b}{2} & h& -\frac{J}{2}&0&\cdots&\vdots\\
  0 & -\frac{J}{2}& h&-\frac{J}{2}&\cdots&\vdots\\
  \vdots & \vdots & \vdots&\ddots&\vdots&\vdots\\
  \vdots & \vdots & \vdots& \vdots&h&-\frac{h_b}{2}\\
  0 & \cdots& \cdots &\cdots  &-\frac{h_b}{2}&0\\
\end{pmatrix},
\end{eqnarray}
and 
\begin{eqnarray}
B&=&
\begin{pmatrix}
   0 & -\frac{h_b}{2}&0&\cdots& \cdots&0\\
   \frac{h_b}{2} & 0& -\frac{J}{2}&0&\cdots&\vdots\\
  0 & \frac{J}{2}& 0&-\frac{J}{2}&\cdots&\vdots\\
  \vdots & \vdots & \vdots&\ddots&\vdots&\vdots\\
  \vdots & \vdots & \vdots& \vdots&0&-\frac{h_b}{2}\\
  0 & \cdots& \cdots &\cdots  &\frac{h_b}{2}&0\\
\end{pmatrix}.
\label{TFIM_EF_B}
\end{eqnarray}
Since the BdG matrix $H_{\rm BdG}$ in Eq.~(\ref{TFIM_BDG_MAT}) can be diagonalized by a unitary matrix $U$, the Hamiltonian~(\ref{TFIM_EF_MAT}) can be written in the diagonal form as 
\begin{eqnarray}
H_{\rm EF}&=&
\begin{pmatrix}
   c^\dagger & c 
\end{pmatrix}
U^\dagger U
\begin{pmatrix}
   A & B\\
   -B & -A
\end{pmatrix}
U^\dagger U
\begin{pmatrix}
   c \\
   c^\dagger
\end{pmatrix}\nonumber\\
&=&
\begin{pmatrix}
   \eta^\dagger & \eta 
\end{pmatrix}
\begin{pmatrix}
   \Lambda & 0\\
   0 & -\Lambda
\end{pmatrix}
\begin{pmatrix}
   \eta \\
   \eta^\dagger
\end{pmatrix}=\sum_{n}\lambda_n(\eta_n^\dagger \eta_n-\eta_n\eta_n^\dagger),\nonumber\\
\label{U}
\end{eqnarray}
where
\begin{eqnarray}
{U}=
\begin{pmatrix}
   l & m\\
   m^* & l^*
\end{pmatrix}
\quad {\rm and}\quad
\begin{pmatrix}
   \eta \\
   \eta^\dagger 
\end{pmatrix}
=
U
\begin{pmatrix}
   c \\
   c^\dagger 
\end{pmatrix}.
\end{eqnarray}
Since the Hamiltonian~(\ref{TFIM_EF_MAT}) possesses a $Z_2$ symmetry: $[H_{\rm EF},\tilde{Z}_2]=0$, where $\tilde{Z}_2=\prod_{i=0}^{L+1}Z_i=\prod_{i=0}^{L+1}(1-2n_i)$, its many-body eigenstates can be chosen to have definite $Z_2$ parity.
In addition, the single-particle eigenvalues $\lambda_n$ of the BdG matrix appear in pairs $(\lambda_n,-\lambda_n)$ due to particle-hole symmetry.
The vacuum state $|0\rangle$ of $H_{\rm EF}$ satisfies
\begin{eqnarray}
   \eta_n|0\rangle=0 \quad {\rm for\ all}\ n.
   \label{vaccum}
\end{eqnarray}
We also note that the zero-energy mode discussed below leads to a degeneracy of the many-body ground state.

\subsubsection{Zero-mode eigenstates and constraints on eigenstates}
The BdG matrix in~(\ref{TFIM_BDG_MAT}) has at least two zero-energy eigenvectors regardless of the values of $J$, $h$, and $h_b$.
Due to the zero-mode degeneracy, there remains freedom in choosing the basis within the degenerate subspace.
For simplicity, we adopt the following convention:
\begin{eqnarray}
   \eta_0=\frac{1}{2}(c_0+c_{L+1}+c_0^\dagger-c_{L+1}^\dagger)
\end{eqnarray}
and
\begin{eqnarray}
   \eta_0^\dagger=\frac{1}{2}(c_0-c_{L+1}+c_0^\dagger+c_{L+1}^\dagger).
\end{eqnarray}
The orthonormality of the Bogoliubov transformation imposes constraints on the remaining eigenstates.
In particular, the elements of $l$ and $m$ must satisfy
\begin{eqnarray}
   l_{n,L+1}=m_{n,L+1}\ {\rm and}\ l_{n,0}=-m_{n,0}\ {\rm for}\ n\neq0.
\end{eqnarray}
Consequently, $\eta_{n\neq0}$ can be written as
\begin{eqnarray}
   \eta_{n}=l_{n,0}(c_0-c_0^\dagger)&+&\sum_{i=1}^{L}(l_{n,i}c_i+m_{n,i}c_i^\dagger)\nonumber\\
   &+&l_{n,L+1}(c_{L+1}+c_{L+1}^\dagger).
\end{eqnarray}
Similarly, the original annihilation operators can be written in terms of the Bogoliubov modes as
\begin{eqnarray}
   c_{i}&=&\sum_{n=1}^{L+1}(l_{n,i}^* \eta_n+m_{n,i}\eta_n^\dagger),\nonumber\\
   c_{0}&=&\frac{1}{2}(\eta_0+\eta_0^\dagger)+\sum_{n=1}^{L+1}(l_{n,0}^* \eta_n-l_{n,0}\eta_n^\dagger),\nonumber\\
   c_{L+1}&=&\frac{1}{2}(\eta_0-\eta_0^\dagger)+\sum_{n=1}^{L+1}(l_{n,L+1}^* \eta_n+l_{n,L+1}\eta_n^\dagger).\nonumber\\
\label{annihilation and creation}
\end{eqnarray}

\subsubsection{Ancilla-spins and parity-broken states}
We now explain how to select definite sectors of the ancilla spins in the fermionic representation.
To this end, it is useful to examine the (anti-)commutation relations involving the ancilla spins~\cite{Campostrini2015,Xavier2020,Jafarizadeh2022}.
We have
\begin{eqnarray}
X_0=c_0+c_0^\dagger=\eta_0+\eta_0^\dagger
\label{Xzero}
\end{eqnarray}
and
\begin{eqnarray}
X_{L+1}=\prod_{i=0}^{L}(1-2n_i)[c_{L+1}+c_{L+1}^\dagger].
\end{eqnarray}
The relevant commutation relations are
\begin{eqnarray}
[X_{0},X_{L+1}]=[X_{L+1},\eta_n]=[X_{L+1},\eta_n^\dagger]=0\ {\rm for\ n\neq 0}.
\label{commutation}
\end{eqnarray}
Using Eq.~(\ref{Xzero}), it is straightforward to obtain the anticommution relations
\begin{eqnarray}
\{X_{0},\eta_n^\dagger \}=\{X_{0},\eta_n \}=0\ {\rm for\ n\neq 0}.
\label{anti-commutation}
\end{eqnarray}
It then follows from Eqs.~(\ref{commutation}) and (\ref{anti-commutation}) that
\begin{eqnarray}
[X_{L+1},\eta_n^\dagger \eta_n]=[X_{0},\eta_n^\dagger \eta_n]=0\ {\rm for\ n\neq 0}.
\label{commutation2}
\end{eqnarray}

We now construct states in definite ancilla-spin sectors.
First, note that the Bogoliubov vacuum $|0\rangle$ has a definite $Z_2$ parity.
Consequently, it cannot be an eigenstate of $X_0$ and $X_{L+1}$, which are $Z_2$-odd operators.
Therefore, we need a state that explicitly breaks $Z_2$ symmetry.
Such ground states can be constructed by taking a linear combination within the degenerate subspace associated with the zero mode:
\begin{eqnarray}
|G_{\pm}\rangle=\frac{1}{\sqrt{2}}(1\pm\eta_0^\dagger)|0\rangle.
\label{Parity-Broken GS}
\end{eqnarray}
The ground states~(\ref{Parity-Broken GS}), often referred to as parity-broken states, are simultaneous eigenstates of both $X_0$ and $X_{L+1}$. 
It is relatively straightforward to show that
\begin{eqnarray}
X_0|G_{\pm}\rangle=\pm|G_{\pm}\rangle.
\end{eqnarray}
Furthermore, Eq.~(\ref{commutation2}) indicates that $X_{L+1}$ does not create any bulk excitations.
Since $X_{L+1}$ commutes with $X_{0}$ [see Eq.~(\ref{commutation})], $X_{L+1}$ preserves the symmetry sector of $X_0$.
Therefore, within the ground-state subspace considered here, $X_{L+1}|G_{\pm}\rangle$ must be proportional to $ |G_{\pm}\rangle$.
Using $X_{L+1}^2=1$, we obtain
\begin{eqnarray}
X_{L+1}|G_{\pm}\rangle=\delta_{L+1}^{\pm}|G_{\pm}\rangle,
\end{eqnarray}
where $(\delta^{\pm}_{L+1})^2=1$. 
The above discussion indicates that, in order to select a desired ancilla-spin sector, we must choose either $|G_+\rangle$ or $|G_-\rangle$ and determine the corresponding value of $\delta^{\pm}_{L+1}$. 
Although the parity-broken states~(\ref{Parity-Broken GS}) are not Gaussian, $\delta^{\pm}_{L+1}$ can be evaluated using generalized Wick’s theorem~\cite{Balian1969} [for details, see Appendix~\ref{Append_A} and the following subsection].

For all parameter values considered in this work, we find that $\delta_{L+1}^+=1$.
Therefore, in what follows, we focus on this case, in which $|G_+\rangle$ belongs to the sector $(\delta_0,\delta_{L+1})=(1,1)$ and corresponds to the ground state of the Hamiltonian~(\ref{TFIM_B}).
If $\delta_{L+1}^+=-1$ and one wants to evaluate quantities in the sector $(\delta_0,\delta_{L+1})=(1,1)$, this sector can be accessed after an appropriate modification of the state.
The details of this discussion are given in Refs.~\cite{Campostrini2015,Xavier2020,Jafarizadeh2022,Hu2023}.

\subsection{Evaluation of correlation functions}
In this subsection, we summarize the correlation functions needed to evaluate quantities such as magnetization and R\'{e}nyi-2 EA.

For two-point functions, we have 
\begin{eqnarray}
C_{i,j}=\langle 0|c_i^\dagger c_j|0\rangle=(m^\dagger m)_{i,j},
\end{eqnarray}
and
\begin{eqnarray}
F_{i,j}=\langle 0|c_i^\dagger c_j^\dagger|0\rangle=(m^\dagger l)_{i,j}.
\end{eqnarray}
For any parity-even operator $\mathcal{O}$ satisfying $[X_0,O]=0$, we have 
\begin{eqnarray}
\langle G_+|\mathcal{O}|G_+\rangle=\langle 0|\mathcal{O}|0\rangle.
\end{eqnarray}
This follows from
\begin{eqnarray}
|G_+\rangle={1+X_0\over\sqrt{2}}|0\rangle,
\end{eqnarray}
together with the fact that the parity-odd terms vanish.
Therefore,
\begin{eqnarray}
\langle G_+|X_i|G_+\rangle=\langle 0|X_0X_i|0\rangle,
\label{X_jcalc}
\end{eqnarray}
and
\begin{eqnarray}
\delta^+_{L+1}=\langle 0|X_0X_{L+1}|0\rangle.
\end{eqnarray}
These quantities can be evaluated using Wick’s theorem.

\section{Magnetization and entanglement asymmetry}
\label{SecIII}
In this section, we focus on the quantities that become finite in the presence of the symmetry-breaking boundary magnetic fields.
Since these quantities vanish when symmetry is preserved and can acquire finite values once the symmetry is broken, they serve as useful diagnostics of symmetry breaking.
We first focus on the local $x$-component of magnetization, denoted by $\langle X_i\rangle=\langle G_+|X_i|G_+\rangle$, and then turn to EA. 
In the following,
we assume $h=1$ and $h_b\ge0$.

\subsection{Magnetization}

\begin{figure}[t]
\begin{center} 
\vspace{0.5cm}
\includegraphics[width=7cm]{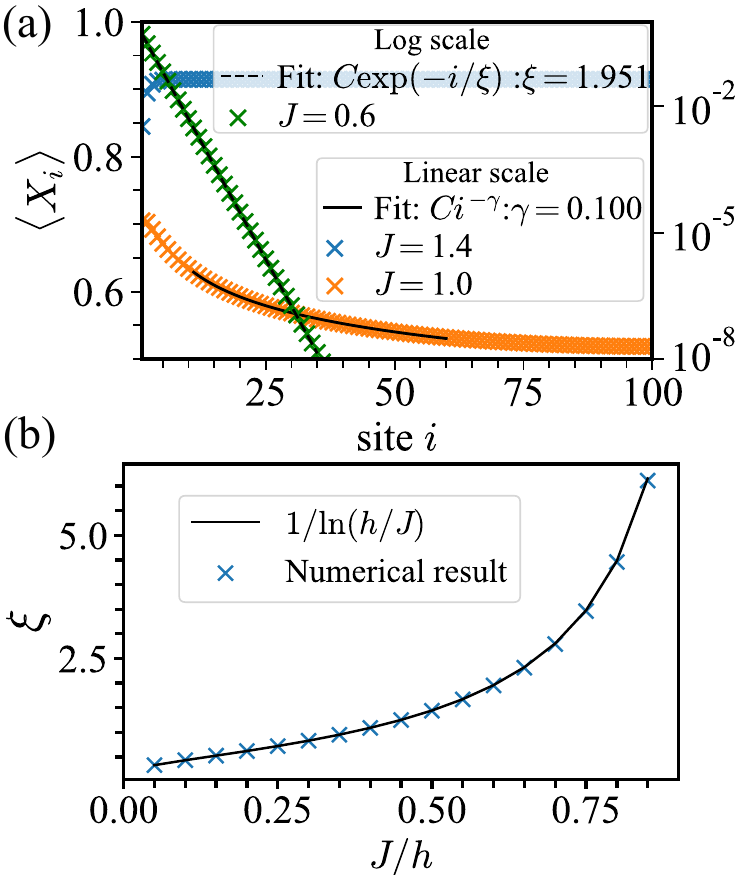}  
\end{center} 
\caption{(a) Expectation value of $X_i$ as a function of the site index $i$ for several values of $J/h$.
The data for $J/h=1$ and $J/h=1.4$ are plotted on a linear scale using the left vertical axis, whereas the data for $J/h=0.6$ are plotted on a logarithmic scale using the right vertical axis.
We set $h_b=0.5$, $h=1$, and $L=200$.
(b) The dependence of $\xi$ as a function of $J/h$.
The numerical data are fitted to the function: $f(i)=C\exp(-\frac{i}{\xi})$.
The solid line represents the perturbative result $\xi=1/\ln(h/J)$.
We set $h_b=0.5$, $h=1$, and $L=100$.
}
\label{Fig_1}
\end{figure}

Using the ancilla spin, the local $x$-component of magnetization can be expressed as a two-point $XX$ correlation function~\cite{Campostrini2015,Xavier2020,Jafarizadeh2022,Hu2023}:
\begin{eqnarray}
\langle G_+|X_i|G_+\rangle=\langle G_+|X_0 X_i|G_+\rangle,
\label{expt_X_j}
\end{eqnarray}
because $|G_+\rangle$ is an eigenstate of $X_0$ with eigenvalue $+1$.
Equation~(\ref{expt_X_j}) can then be evaluated using Wick's theorem and Eq.~(\ref{X_jcalc}).

Figure~\ref{Fig_1} shows $\langle X_i\rangle$ as a function of the site index $i$ for several values of $J/h$. 
For $J/h<1$, we observe that $\langle X_i\rangle$ decays exponentially with the distance from the boundary, which is well captured by an exponential fit to the numerical data.
To obtain a better understanding of this behavior in the limit $J\ll h$, we perform a perturbative analysis.
We find that the local $x$-component exhibits exponential decay $\langle X_i\rangle\propto \exp(-i/\xi)$, where $\xi=1/\ln(h/J)$, in agreement with the numerical result.
Details of the calculation are provided in Appendix~\ref{Append_B}.
As shown in Fig.~\ref{Fig_1}(b), $\xi$ extracted from the numerical fitting agrees well with the perturbative prediction $\xi=1/\ln(h/J)$.
Intuitively, this behavior can be understood as the boundary perturbation penetrating into the bulk over a distance of the order of the correlation length.

At the critical point $J/h=1$, $\langle X_i\rangle$ exhibits an algebraic decay as a function of $i$.
Although symmetry-breaking magnetic fields are applied at both ends in our system, the influence of the right boundary is expected to become negligible near the left boundary in the thermodynamic limit.
For a semi-infinite chain with a magnetic field applied at one boundary, the magnetization profile is governed by the boundary length scale $\xi_b\sim h_b^{-2}$.
At distances $i\gg \xi_b$, the magnetization takes the asymptotic scaling form $\langle X_i\rangle\propto i^{-1/8}$ under the fixed boundary condition~\cite{CHATTERJEE1994,Sela2012}.
We find that $\gamma=0.100$, extracted from the numerical result, is qualitatively consistent with the asymptotic behavior $\gamma=1/8=0.125$.
The remaining deviation may come from finite-size effects, the finite crossover scale $\xi_b$, and the presence of the right boundary magnetic field.

In the ordered phase $J/h=1.4$, $\langle X_i\rangle$ approaches the exact spontaneous magnetization of the TFIM $\langle X_i\rangle\sim(1-(h/J)^2)^{1/8}\sim0.91$~\cite{PFEUTY197079}, rather than decaying to zero.
The behavior of both the bulk and boundary magnetization can be understood using perturbative analysis.
Details of the perturbative analysis are provided in Appendix~\ref{Append_B}.

\subsection{Definition of entanglement asymmetry}
We have so far observed that a boundary magnetic field induces $Z_2$ symmetry-breaking in a manner that depends on the phase of the TFIM.  
As a complementary analysis, we now quantify the extent of symmetry breaking using an information-theoretic measure known as EA.
In this subsection, we introduce the definition of EA and describe how it is evaluated in the present setup.

EA measures the degree of symmetry breaking within a subsystem.
To define it, we decompose the whole system into two complementary subsystems $A$ and $B$ [see Fig.~\ref{Fig_2}(a)].
The physical state of subsystem $A$ is completely characterized by the reduced density matrix:
\begin{eqnarray}
\rho_A={\rm Tr}_B[|\Psi\rangle\langle\Psi|].
\label{def_rho_A}
\end{eqnarray}
Then, EA is defined by comparing $\rho_A$ with its symmetrized counterpart.
In particular, the EA associated with $Z_2$ symmetry breaking is defined as 
\begin{eqnarray}
\Delta S_A(\ell_A)&=&{\rm Tr_A}[\rho_A\ln\rho_A-\bar{\rho}_A\ln\bar{\rho}_A],
\label{def_EA}
\end{eqnarray}
where $\bar{\rho}_A=\sum_{\alpha=\pm}\Pi_\alpha\rho_A\Pi_\alpha$, $\Pi_\alpha=(I+\alpha U_{Z_2})/2$, and  $U_{Z_2}=\prod_{i=1}^{\ell_A}Z_i$.
Here, $\ell_A$ denotes the site index specifying the right boundary of subsystem $A$.
We note that the ancilla spin at site $0$ is an auxiliary degree of freedom introduced for the fermionic calculation and is therefore not included in the physical symmetry operation $U_{Z_2}$.
For later use, we also define the EA for subsystem $B$ as 
\begin{eqnarray}
\Delta S_B(\ell_B)&=&{\rm Tr_B}[\rho_B\ln\rho_B-\bar{\rho}_B\ln\bar{\rho}_B],
\label{def_EA_B}
\end{eqnarray}
where
$\rho_B={\rm Tr}_A[|\Psi\rangle\langle\Psi|]$, $\bar{\rho}_B=\sum_{\beta=\pm}\Pi_\beta\rho_B\Pi_\beta$, $\Pi_\beta=(I+\beta U_{Z_2}^B)/2$, and  $U_{Z_2}^B=\prod_{i=\ell_B}^{L}Z_i$.
Here, $\ell_B$ specifies the left boundary of subsystem $B$.
In the following, we mainly focus on subsystem $A$.

EA has the following two important properties:
(i) $\Delta S_A\ge0$, since the definition of EA is nothing but the relative entropy between $\rho_A$ and $\bar{\rho}_A$, and
(ii) $\Delta S_A=0$ if and only if the reduced density matrix is symmetric under the subsystem symmetry operation. 
Although the EA (\ref{def_EA}) provides a concrete information-theoretic measure of symmetry breaking, its direct calculation is generally challenging because it requires the full spectrum of the reduced density matrix.
Thus, many studies instead focus on R\'enyi generalizations of EA.
Specifically, the R\'enyi-2 EA $\Delta S_2(\ell_A)$ given by 
\begin{eqnarray}
\Delta S_2(\ell_A)&=&\ln{\rm Tr_A}[\rho_A^2]-\ln{\rm Tr_A}[\bar{\rho}_A^2]\nonumber\\
&=&-\ln[{1\over2}(1+{{\rm Tr_A}[{\rho}_AU_{Z_2}{\rho}_AU_{Z_2}]\over{\rm Tr_A}[{\rho}_A^2]})],
\label{def_REA}
\end{eqnarray}
has attracted attention because it can be efficiently computed for Gaussian states~\cite{Ares2023} and is experimentally accessible using randomized measurements~\cite{Joshi2024}. 
While the R\'enyi-2 EA is not generally equivalent to a quantum relative entropy and does not necessarily satisfy certain information-theoretic properties~\cite{kusuki2026}, it still satisfies properties (i) and (ii).
\subsection{Efficient calculation of entanglement asymmetry for parity-broken states}
While efficient numerical methods for calculating $\Delta S_2$ for Gaussian states have been developed, the parity-broken state considered here is non-Gaussian, requiring an extension of these techniques.
To this end, we first need to determine the reduced density matrix associated with a parity-broken state.
Although a more general derivation of such a reduced density matrix is provided in Refs.~\cite{Xavier2020,Jafarizadeh2022}, we present a simplified derivation adapted to the present setup.

First, note that $\rho_A$ can be written as 
\begin{eqnarray}
\rho_A&=&{1\over2}{\rm Tr_B}[(1+\eta_0^\dagger)|0\rangle\langle0|(1+\eta_0)]\nonumber\\
&=&{1\over2}{\rm Tr_B}[(1+X_0)|0\rangle\langle0|(1+X_0)]\nonumber\\
&=&{1\over2}(1+X_0)\rho_A^0(1+X_0),
\label{derivation_rho_A1}
\end{eqnarray}
where $\rho_A^0={\rm Tr_B}[|0\rangle\langle0|]$.
In the second line of Eq.~(\ref{derivation_rho_A1}), we use $\eta_0|0\rangle=0$.
Since $|0\rangle$ is a fermionic Gaussian state, $\rho_A^0$ is also Gaussian and is therefore completely determined by the two-point correlation functions restricted to subsystem $A$.
To this end, we introduce Majorana operators as 
\begin{eqnarray}
\Phi=
\begin{pmatrix}
   {a}_0 \\ 
   \vdots \\
   {a}_{\ell_A}\\
   {a}_{\ell_A+1} \\ 
   \vdots \\
   {a}_{2\ell_A+1}
\end{pmatrix}
=W
\begin{pmatrix}
   c_0 \\ 
   \vdots \\
   c_{\ell_A}\\
   c_0^\dagger \\ 
   \vdots \\
   c_{\ell_A}^\dagger
\end{pmatrix},
\label{majorana}
\end{eqnarray}
where
\begin{eqnarray}
W=
\begin{pmatrix}
   I & I\\
   -iI & iI
\end{pmatrix}.
\label{W_matrix}
\end{eqnarray}
Then, we define the restricted covariance matrix $M$ as
\begin{eqnarray}
M_{i,i'}=\langle {a}_i {a}_{i'} \rangle=\delta_{i,i'}+i\Gamma_{i,i'},
\end{eqnarray}
where $\Gamma^T=-\Gamma$.
Following the well-established Peschel formula~\cite{Peschel2009,Fagotti2010}, $\rho_A^0$ can be written as
\begin{eqnarray}
\rho_A^0&=&\det ({I+i\Gamma\over2})^{1/2}\exp[{1\over4}\Phi^T \ln({I-i\Gamma\over I+i\Gamma})\Phi].
\label{Correlation-matrix3}
\end{eqnarray}
Although the calculation is somewhat involved, we can show 
\begin{eqnarray}
[\rho_A^0,X_0]=0.
\label{commutation_rho_0}
\end{eqnarray}
The detailed derivation of Eq.~(\ref{commutation_rho_0}) is provided in Appendix~\ref{Append_C}.
We then obtain
\begin{eqnarray}
\rho_A=(1+X_0)\rho_A^0, 
\label{derivation_rho_A3}
\end{eqnarray}
which is consistent with previous studies~\cite{Xavier2020,Jafarizadeh2022}.

We next consider the calculation of EA.
Substituting Eq.~(\ref{derivation_rho_A3}) into Eq.~(\ref{def_REA}), 
$\Delta S_2$ can be written as follows:
\begin{eqnarray}
\Delta S_2=-\ln[{1\over2}(1+{\mathcal{Z}_2})]
\label{def_REA1}
\end{eqnarray}
where  ${\mathcal Z}_2={\rm Tr_A}[(1+X_0){\rho}_A^0{\tilde{\rho}}_A^0]/{\rm Tr_A}[(1+X_0){\rho}_A^0{\rho}_A^0]$ and $\tilde{\rho}_A^{0}=X_0U_{Z_2}\rho_A^0U_{Z_2}X_0$.
Notably, Eq.~(\ref{def_REA1}) can be simplified.
The key observation is that the unitary operator $X_0U_{Z_2}$ only flips the sign of a single Majorana operator:
\begin{eqnarray}
X_0U_{Z_2}a_{\ell_A+1}U_{Z_2}X_0=-a_{\ell_A+1},
\end{eqnarray}
while leaving all other Majorana operators unchanged.
Since this transformation only changes the sign of a Majorana operator, 
$\tilde{\rho}_A^{0}$ remains Gaussian and can be written as
\begin{eqnarray}
\tilde{\rho}_A^{0}=\det ({I+i\Gamma'\over2})^{1/2}\exp[{1\over4}\Phi^T \ln({I-i\Gamma'\over I+i\Gamma'})\Phi].\nonumber\\
\label{Correlation-matrix4}
\end{eqnarray}
Here, $\Gamma'=R\Gamma R^T$ and $R=I-2e_{\ell_A+1}e_{\ell_A+1}^T$, where $I$ is the identity matrix and $e_{\ell_A+1}$ denotes the unit vector in the $(\ell_A+1)$-th direction.
The product of two fermionic Gaussian density operators remains Gaussian and is parity even; therefore, the expectation value of parity-odd operators such as $X_0$ vanishes.
Using the Gaussian overlap formula~\cite{Bravyi2005}, 
we now obtain a form of Eq.~(\ref{def_REA1}) that is more convenient for numerical calculation:
\begin{eqnarray}
\Delta S_2&=&-\ln[{1\over2}(1+{\sqrt{\det [I-\Gamma\Gamma']} \over \sqrt{\det[I-\Gamma^2]}})].
\label{def_REA2}
\end{eqnarray}
Notably, the matrix determinant lemma~\cite{Harville1997} allows us to 
further simplify Eq.~(\ref{def_REA2}) as
\begin{eqnarray}
\Delta S_2&=&-\ln[(I-\Gamma^2)^{-1}_{\ell_A+1,\ell_A+1}].
\label{def_REA3}
\end{eqnarray}
The details of the derivation of Eq.~(\ref{def_REA3}) are provided in Appendix~\ref{Append_D}.

For later discussion, we derive the following useful inequalities:
\begin{eqnarray}
-\ln(1-{1\over2}s)\le\Delta S_2\le\ln(1+s),
\label{def_inequality1}
\end{eqnarray}
where $s=\sum_{i=0}^{2\ell_A+1}|\Gamma_{\ell_A+1,i}|^2$,
and
\begin{eqnarray}
\Delta S_2&\ge&-\ln(1-{1\over2}\underset{1\le i\le\ell_A}{\rm max}\langle0| X_0X_i|0\rangle^2)\nonumber\\
&=&-\ln(1-{1\over2}\underset{1\le i\le\ell_A}{\rm max}\langle G_+| X_i|G_+\rangle^2)
\label{def_inequality2}
\end{eqnarray}
The detailed derivations are provided in Appendix~\ref{Append_E}.

\subsection{Entanglement asymmetry}
\begin{figure}[t]
\begin{center} 
\vspace{0.5cm}
\includegraphics[width=7cm]{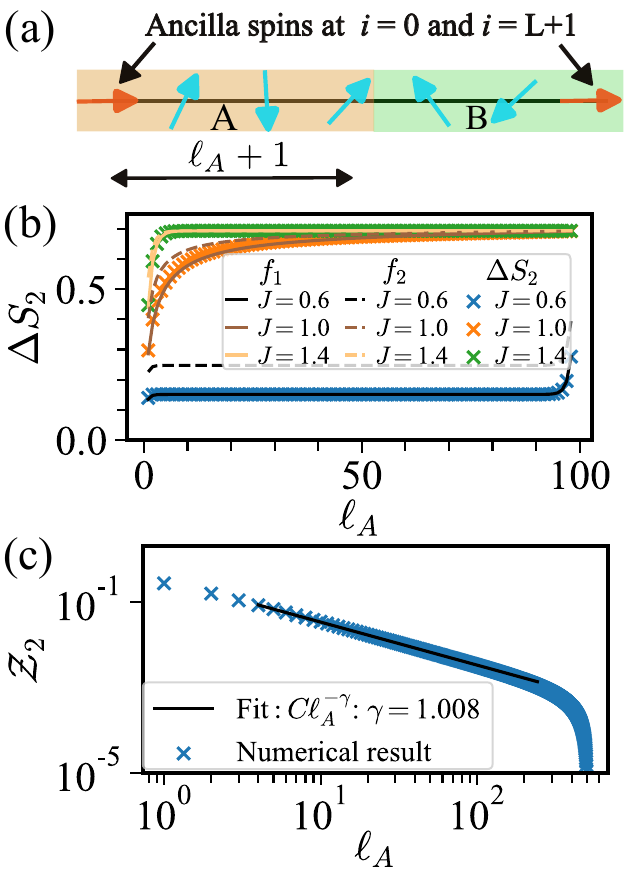}  
\end{center} 
\caption{(a) Schematic illustration of the bipartition of the system.
(b) Numerical results for $\Delta S_2(\ell_A)$ as a function of the subsystem index $\ell_A$ for several values of $J/h$. 
We set $h_b=0.5$, $h=1$, and $L=100$.
(c) Numerical result for ${\mathcal Z}_2$ as a function of the subsystem index $\ell_A$ for $J/h=1$.
We set $h_b=0.5$, $h=1$, and $L=500$.
The solid line represents a fit to the power-law function $f(\ell_A)=C\ell_A^{-\gamma}$.
}
\label{Fig_2}
\end{figure}
Having derived the analytical expression for $\Delta S_2$, we now examine its behavior numerically.
In particular, we investigate how $\Delta S_2$ depends on the subsystem index $\ell_A$ in each phase.

Figure~\ref{Fig_2}(b) shows $\Delta S_2$ for several values of $J/h$, represented by symbols, together with the corresponding lower and upper bounds, $f_1=-\ln(1-{1\over2}s)$ and $f_2=\ln(1+s)$, shown by solid and dashed lines, respectively.
For $J/h<1$, as $\ell_A$ increases, $\Delta S_2$ initially increases, reaches a plateau in an intermediate regime of $\ell_A$, and then increases again as $\ell_A$ approaches the right boundary.
This behavior appears to be qualitatively consistent with that of $\langle X_i\rangle$, particularly the plateau region in the bulk.  
Since the magnetization---an indicator of symmetry breaking---becomes exponentially small in the bulk, enlarging the subsystem further into the bulk adds little information about the broken symmetry.
By contrast, as $\ell_A$ approaches $L$, the subsystem includes sites with large magnetization, resulting in an increase in $\Delta S_2$.

At the critical point of the TFIM, $J/h=1$, as 
$\ell_A$ increases, $\Delta S_2$ tends to increase and gradually approaches $\ln(2)$, without showing the plateau observed for $J/h<1$.
Equivalently, ${\mathcal Z}_2=\sqrt{{\det [I-\Gamma\Gamma']}/{\det[I-\Gamma^2]}}$ decreases with increasing $\ell_A$.
Although our setup differs from that considered in the previous study~\cite{Fossati2025}, where an algebraic decay of  ${\mathcal Z}_2\propto \ell_A^{-1}$ was obtained,
our data [see Fig.~\ref{Fig_2}(c)] are also compatible with an algebraic decay. 
The extracted exponent is relatively close to the previously reported exponent $\gamma=1$.
The agreement between the numerical results and the BCFT prediction suggests that, even for finite boundary magnetic fields, the boundary condition flows toward the symmetry-breaking conformal boundary fixed point for large $\ell_A$.

For $J/h>1$, $\Delta S_2$ exhibits a sharp increase and remains close to $\ln(2)$, with only a weak dependence on $\ell_A$.
Intuitively, this behavior may be understood in terms of the characteristics of $\Delta S_2$.
Once the subsystem contains enough information to identify the symmetry-broken state, for example, through a finite magnetization, further enlarging the subsystem into a region where the magnetization remains finite provides little additional information.
This intuitive explanation is consistent with Eq.~(\ref{def_inequality2}); $\Delta S_2$ is lower-bounded by the maximum squared local magnetization within the subsystem rather than by the sum or average magnetization.

Finally, we note that both the upper and lower bounds capture the qualitative behavior of $\Delta S_2$.
We find that the lower bound $f_1$ closely agrees with $\Delta S_2$ over the parameter range considered here.
This indicates that the lower bound $f_1$ is tight and accurately captures not only the qualitative behavior but also the magnitude of $\Delta S_2$.
Although $f_2$ is a somewhat loose upper bound for $\Delta S_2$ when $J/h<1$, it is still qualitatively consistent with $\Delta S_2$, and the bound becomes tighter when $J/h\ge1$.  
Interestingly, the quantity $s$, which determines both the upper and lower bounds of $\Delta S_2$, can be written as a sum of squared correlators: 
\begin{eqnarray}
s=\sum_{i=1}^{\ell_A} \langle X_0Z_1\cdots Z_{i-1}X_i\rangle^2+\sum_{i=1}^{\ell_A} \langle X_0Z_1\cdots Z_{i-1}Y_i\rangle^2.
\end{eqnarray}
These correlators can become finite in the presence of nonzero boundary fields.
Thus, roughly speaking, Eq.~(\ref{def_inequality1}) relates $\Delta S_2$ to nonlocal spin correlation functions that reflect the effects of symmetry-breaking perturbations.

\section{Quench dynamics}
Having established the ground-state properties of the system, we now focus on its nonequilibrium dynamics.
We first explain the initial state and the quench protocol.
We then numerically investigate the quench dynamics induced by boundary magnetic fields.

\begin{figure*}[t]
\begin{center} 
\vspace{0.5cm}
\includegraphics[width=17cm]{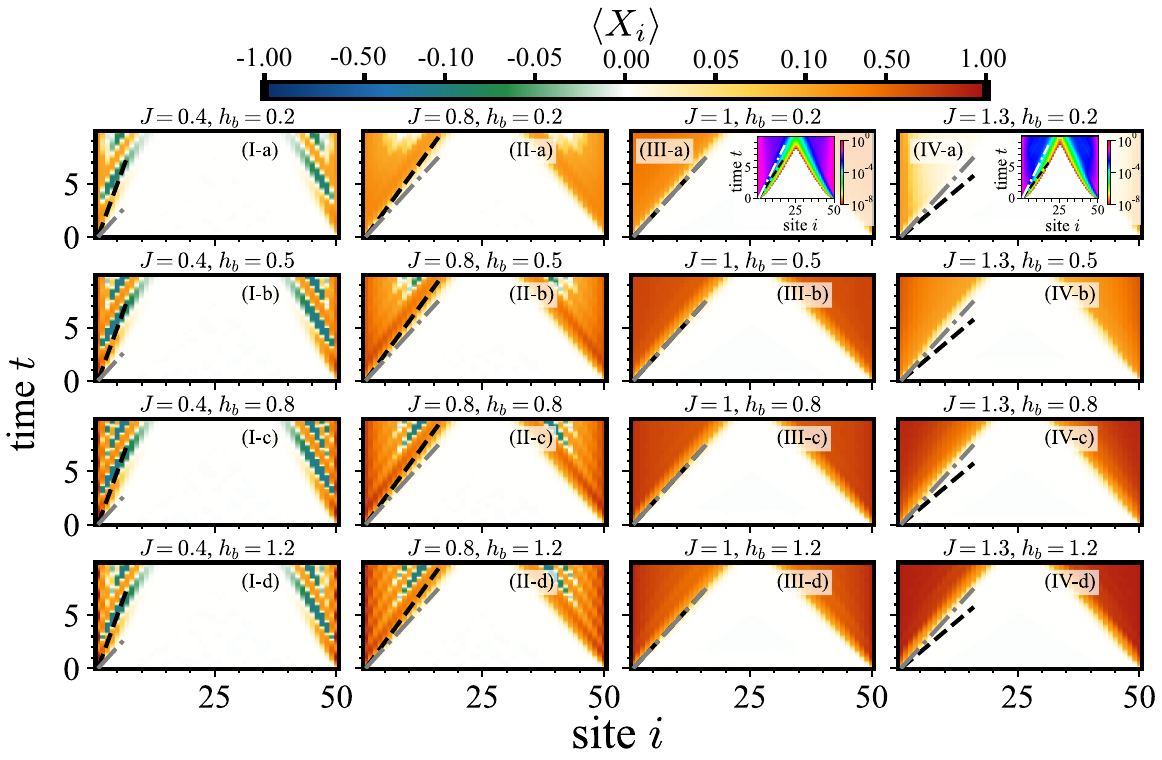}  
\end{center} 
\caption{
Color maps of $\langle X_i\rangle$ as a function of the site index $i$ and time $t$.
Each row corresponds to a different value of $h_b$ while each column corresponds to a different value of $J/h$.
The black dashed and gray dash-dotted lines indicate the propagation fronts corresponding to the quasiparticle velocity $2J$ and $2h$, respectively.
The values of $h_b$ increase from top to bottom, whereas $J/h$ increases from left to right.
The insets in panels (III-a) and (IV-a) show $|\langle X_i\rangle|$ on a logarithmic scale for the same parameters as in the main panels.
In these insets, the white dash-dotted line indicates the propagation front corresponding to the quasiparticle velocity $2h$.
We set $L=50$.
}
\label{Fig_3}
\end{figure*}

\subsection{Boundary-quench protocol}
In this work, we consider the following quench protocol.
We first prepare the physical chain in the ground state of the TFIM, with the ancilla spins in the $X=+1$ states:
\begin{eqnarray}
   |\Psi(t=0)\rangle=|+\rangle_0\otimes|{\rm GS}(H_0)\rangle\otimes|+\rangle_{L+1},
\end{eqnarray}
where $X_{0}|+\rangle_{0}=|+\rangle_{0}$, $X_{L+1}|+\rangle_{L+1}=|+\rangle_{L+1}$, and $|{\rm GS}(\hat{H}_0)\rangle$ is the even-parity ground state of the TFIM.
At $t=0$, the boundary magnetic fields are suddenly switched on to finite values, and $|\Psi(t=0)\rangle$ evolves according to the post-quench Hamiltonian~(\ref{TFIM_B}).

To implement this boundary quench in the fermionic calculation, 
we choose the initial state $|\Psi(t=0)\rangle$ in the ancilla sector $(\delta_0,\delta_{L+1})=(1,1)$, so that the enlarged system corresponds to the physical boundary fields ($h_b>0$). 
In the following, we denote the Bogoliubov operator associated with the pre-quench Hamiltonian (\ref{TFIM_EF_MAT}) by $\tilde{\eta}_n$ and $\tilde{\eta}_n^\dagger$.
Notably, for the enlarged pre-quench system, the ancilla degrees of freedom are
completely decoupled from the bulk.
In this case, in addition to $\eta_0=\tilde{\eta}_0$ and $\eta_0^\dagger=\tilde{\eta}_0^\dagger$,
the BdG matrix of the pre-quench Hamiltonian (\ref{TFIM_EF_MAT}) has an additional pair of zero-energy eigenvectors.
For simplicity, we adopt the following definition:
\begin{eqnarray}
   \tilde{\eta}_1=\frac{1}{2}(-c_0+c_{L+1}+c_0^\dagger+c_{L+1}^\dagger)
\end{eqnarray}
and
\begin{eqnarray}
   \tilde{\eta}_1^\dagger=\frac{1}{2}(c_0+c_{L+1}-c_0^\dagger+c_{L+1}^\dagger).
\end{eqnarray}
Using these modes, the initial state in the $(\delta_0,\delta_{L+1})=(1,1)$ sector can be constructed in the same manner as the parity-broken state $|G_+\rangle$ defined in Eq.~(\ref{Parity-Broken GS}):
\begin{eqnarray}
|\Psi(t=0)\rangle={1+\tilde{\eta}^\dagger_0\over\sqrt{2}}|\tilde{0}\rangle.
\end{eqnarray}
Here, $|\tilde{0}\rangle$ is the vacuum state of Hamiltonian~(\ref{TFIM0_EF}) and satisfies 
\begin{eqnarray}
   \tilde{\eta}_n|\tilde{0}\rangle=0 \quad {\rm for\ all}\ n.
\end{eqnarray}

The state $|\Psi(t=0)\rangle$ then evolves under the post-quench Hamiltonian~(\ref{TFIM_EF}).
Since $[\hat{H}_{\rm EF},X_0]=[\hat{H}_{\rm EF},X_{L+1}]=0$, the initially selected ancilla sectors remain unchanged throughout the time evolution.

\begin{figure*}[t]
\begin{center} 
\vspace{0.5cm}
\includegraphics[width=17cm]{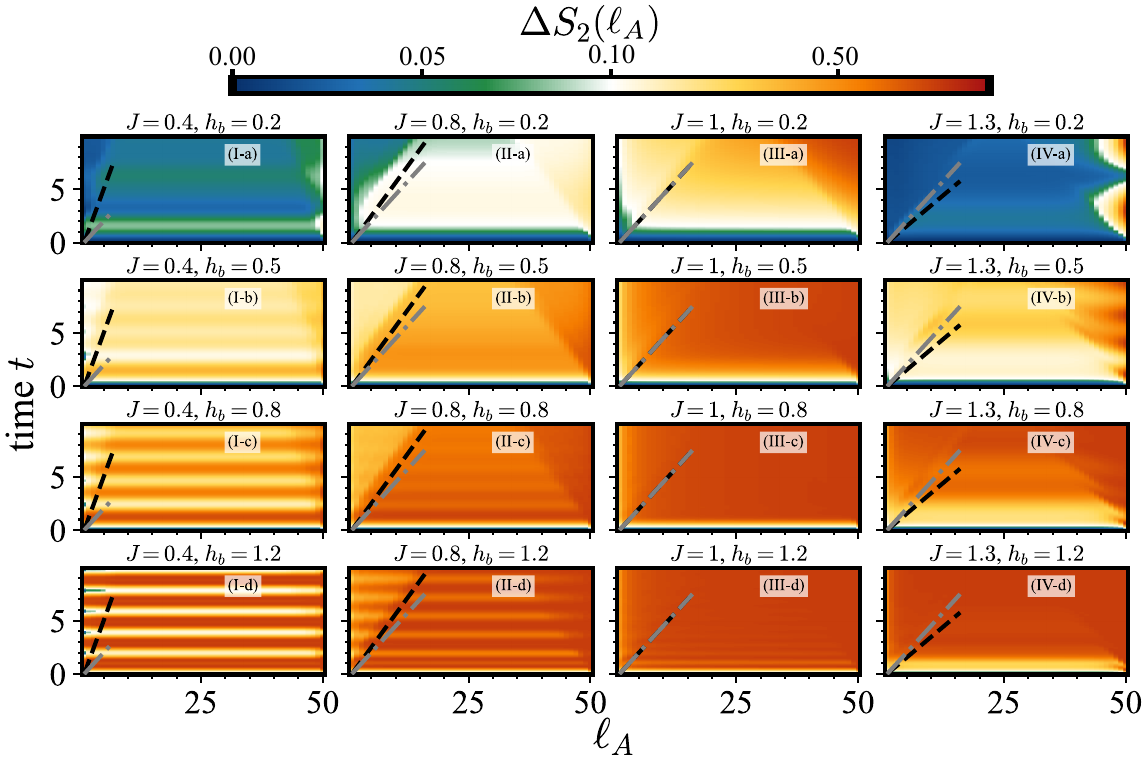}  
\end{center} 
\caption{
Color maps of $\Delta S_2(\ell_A)$ as a function of the
subsystem index $\ell_A$ and time $t$.
Each row corresponds to a different value of $h_b$ while each column corresponds to a different value of $J/h$.
The values of $h_b$ increase from top to bottom, whereas $J/h$ increases from left to right.
We set $L=50$.
The black dashed and gray dash-dotted lines indicate the propagation fronts corresponding to the quasiparticle velocity $2J$ and $2h$, respectively.
}
\label{Fig_4}
\end{figure*}

\subsection{Gaussian formalism for time evolution}
We introduce several useful techniques for analyzing time-dependent magnetization and EA.
With these techniques, the problem can be reduced to the standard quadratic-fermion calculation.

Starting from the initial state specified above,
we evolve the system under the post-quench Hamiltonian~(\ref{TFIM_EF}):
\begin{eqnarray}
   |\Psi(t)\rangle=e^{-i\hat{H}_{\rm EF}t}{1+\tilde{\eta}_0^\dagger\over\sqrt{2}}|\tilde{0}\rangle={1+\tilde{\eta}_0^\dagger\over\sqrt{2}}|\phi(t)\rangle,
\label{time-evolved}
\end{eqnarray}
where $|\phi(t)\rangle=e^{-i\hat{H}_{\rm EF}t}|\tilde{0}\rangle$ and we use $[\hat{H}_{\rm EF},\tilde{\eta}_0^\dagger]=0$.
Here, the state $|\phi(t)\rangle$ remains Gaussian because the initial vacuum is Gaussian and the Hamiltonian is quadratic.
Following the time-dependent Bogoliubov-de Gennes construction~\cite{Mbeng2024,Russomanno2016}, we define the time-dependent annihilation operator associated with the time-evolved Gaussian state as 
\begin{eqnarray}
\tilde{\eta}_{n}(t)=e^{-i\hat{H}_{\rm EF}t}\tilde{\eta}_{n}(t=0)e^{i\hat{H}_{\rm EF}t}.
\label{eta-tevo}
\end{eqnarray}
These operators satisfy $\tilde{\eta}_n(t)|\phi(t)\rangle=0$ for all $n$ and all $t$.
Since $\tilde{\eta}_0(t) =\tilde{\eta}_0=\eta_0$, the orthonormality of the Bogoliubov transformation constrains the remaining time-dependent modes $\tilde{\eta}_{n\neq0}(t)$ in the same way as in the ground-state case.
Consequently, the formalism developed for the ground state remains valid with the replacement $\eta_n\to \tilde{\eta}_n(t)$.
In practice, instead of directly calculating $\tilde{\eta}_n(t)$, we represent the quadratic fermionic Hamiltonian as a ($2L+4) \times( 2L+4$) BdG matrix in Eq.~(\ref{TFIM_BDG_MAT}) and directly evaluate the time-dependent correlation matrix of $|\phi(t)\rangle$: 
\begin{eqnarray}
{\mathbb  G}(t)&=&\langle\tilde{0}|e^{i\hat{H}_{{\rm EF}}t}
\begin{pmatrix}
   cc^\dagger & cc\\
   c^\dagger c^\dagger & c^\dagger c
\end{pmatrix}
e^{-i\hat{H}_{{\rm EF}}t}|\tilde{0}\rangle
\nonumber\\
&=&{\mathbb  U}(t)
  \langle\tilde{0}|
\begin{pmatrix}
cc^\dagger & cc\\
   c^\dagger c^\dagger & c^\dagger c
\end{pmatrix}
|\tilde{0}\rangle
{\mathbb  U}^\dagger(t),
\label{time-dep-corr}
\end{eqnarray}
where ${\mathbb  U}(t)=\exp(-2iH_{\rm BdG}t)$ and $H_{\rm BdG}$ is the ($2L+4) \times (2L+4$) BdG matrix defined in Eq.~(\ref{TFIM_BDG_MAT}).

\begin{figure}[t]
\begin{center} 
\vspace{0.5cm}
\includegraphics[width=8cm]{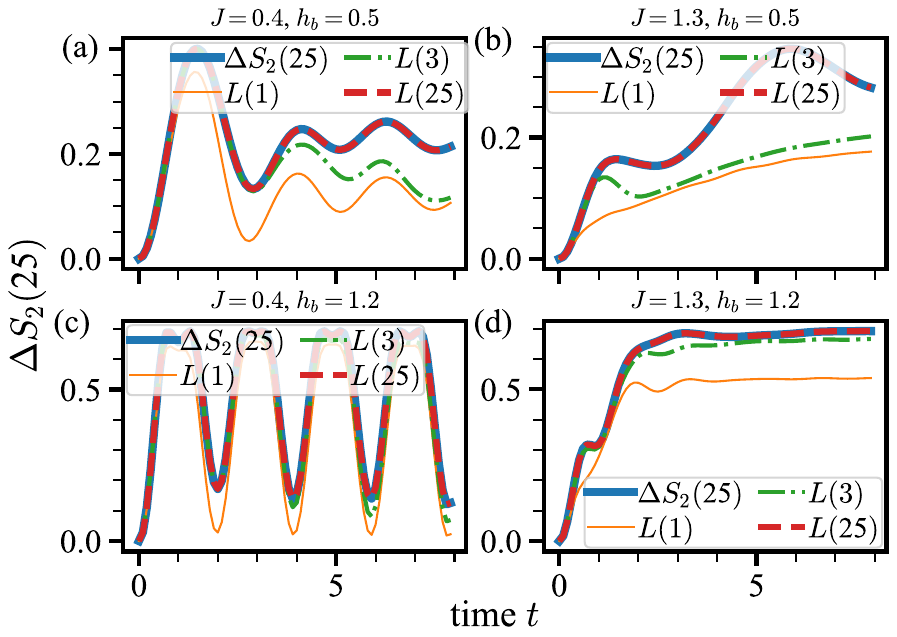}  
\end{center} 
\caption{Time evolution of the $\Delta S_{2}(\ell_A)$ in the bulk and $L(r)$ for several values of $r$ at different $J$ and $h_b$.
Each row corresponds to a different value of $h_b$ while each column corresponds to a different value of $J$.
The values of $h_b$ increase from top to bottom, whereas $J$ increases from left to right.
We set $L=50$.
}
\label{Fig_5}
\end{figure}

\subsection{Propagation of magnetization}

Figure~\ref{Fig_3} shows the time evolution of $\langle X_i\rangle$ after the boundary quench for several values of $J/h$ and $h_b$.
To compare the propagation of the magnetization with the bulk quasiparticle velocity~\cite{Calabrese2006}, we also plot the lines $t=\frac{i-1}{2J}$ and $t=\frac{i-1}{2h}$~\footnote{Since the site index $i$ starts from $i=1$, we take into account the shift by one site. }, shown by the black dashed and gray dash-dotted lines, respectively. 

Overall, the propagation front of the magnetization is visually consistent with one of these lines corresponding to the maximum quasiparticle velocity $v_{\rm max}=2\min(J,h)$.
Notably, around $J/h=1$, the propagation front of magnetization is remarkably consistent with the maximum quasiparticle velocity.

For $J/h<1$, while a small deviation from the lines can be seen, the propagation front of the magnetization seems to be described by the velocity $2J$.
This small deviation may reflect the ambiguity in identifying the front position in the presence of oscillations.

For $J/h>1$, the propagation front of the magnetization is instead described by the velocity $2h$.
When $h_b$ is small, the propagation front becomes difficult to identify directly.
Nevertheless, the corresponding logarithmic plots of the absolute value of the magnetization shown in the insets of Fig.~(\ref{Fig_3}) confirm that a similar propagating structure remains and is approximately consistent with the velocity $2h$.

\subsection{Entanglement asymmetry dynamics}

Figure~\ref{Fig_4} shows the time evolution of $\Delta S_2(\ell_A)$.
To highlight the characteristic behavior, we focus on several representative cases.
First, we note that $\Delta S_2(\ell_A)$ does not exhibit the clear quasiparticle-like propagation observed in the local magnetization.
This behavior is consistent with the fact that its lower bound depends on the maximum local magnetization within the subsystem.
Because of this lower bound, $\Delta S_2(\ell_A)$ can already become large even before the local magnetization has propagated throughout the entire subsystem.

The qualitative argument above is based on the simpler bound in Eq.~(\ref{def_inequality2}).
To examine this interpretation more quantitatively, we introduce a site-resolved form of the tighter lower bound in Eq.~(\ref{def_inequality1}):
\begin{eqnarray}
L(r)=-\ln(1-\frac{1}{2}s(r)),
\end{eqnarray}
where
\begin{eqnarray}
s(r)=\sum_{i=0}^{r}|\Gamma_{\ell_A+1,i}|^2+\sum_{i=\ell_A+1}^{\ell_A+1+r}|\Gamma_{\ell_A+1,i}|^2.
\end{eqnarray}
Since $L(r)\le L(r')$, where $r\le r'$, $L(r)$ reveals how correlations within the subsystem successively contribute to the lower bound of $\Delta S_{2}(\ell_A)$.
If $L(r)$ with $r<<\ell_A$ approaches $\Delta S_{2}(\ell_A)$ already in small $r$, this supports the above interpretation.
Figure~\ref{Fig_5} shows $L(r)$ together with $\Delta S_{2}(\ell_A)$ in the bulk for several representative parameter values, which are discussed in more detail below.

We now examine several representative cases in more detail.
For small $J/h$ [see Fig.~\ref{Fig_4}(I-b) and (I-d)], $\Delta S_2(\ell_A)$ tends to exhibit oscillatory behavior that is almost independent of $\ell_A$. 
To clarify the origin of this behavior, we turn to Fig.~\ref{Fig_5}(a) and (c), where $L(r)$ and $\Delta S_{2}(\ell_A)$ in the bulk are plotted for the same parameter sets.
Interestingly, even for small $r$, $\Delta S_{2}(\ell_A)$ and $L(r)$ exhibit oscillatory behaviors with similar periods and are of the same order of magnitude, suggesting a significant boundary contribution to $\Delta S_2(\ell_A)$.
For large $J/h$ and $h_b$ [see Fig.~\ref{Fig_4}(IV-d)], $\Delta S_2(\ell_A)$ saturates after a short time, and this behavior is nearly independent of $\ell_A$.
As shown in Fig.~\ref{Fig_5}(d), $L(r)$ for small $r$ qualitatively captures this behavior.  
This finding further indicates that $\Delta S_2(\ell_A)$ is strongly influenced by the boundary contribution rather than by the spatial propagation of symmetry breaking.
In the case of large $J/h$ and small $h_b$ [see Fig.~\ref{Fig_4}(IV-b)], $\Delta S_2(\ell_A)$ remains nearly constant in the bulk and exhibits oscillatory behavior over time.
In contrast to the other parameter regimes,
as shown in Fig.~\ref{Fig_5}(b),
the boundary contribution alone --- $L(r)$ for small $r$ --- does not fully capture the oscillatory behavior or the magnitude of $\Delta S_2(\ell_A)$ in the bulk.
However, increasing $r$ improves the agreement between $\Delta S_2(\ell_A)$ and $L(r)$.
This result suggests that $\Delta S_2(\ell_A)$ also reflects the spatially extended nature of symmetry-breaking information.

\subsection{Asymmetry contour}
\begin{figure*}[t]
\begin{center} 
\vspace{0.5cm}
\includegraphics[width=17cm]{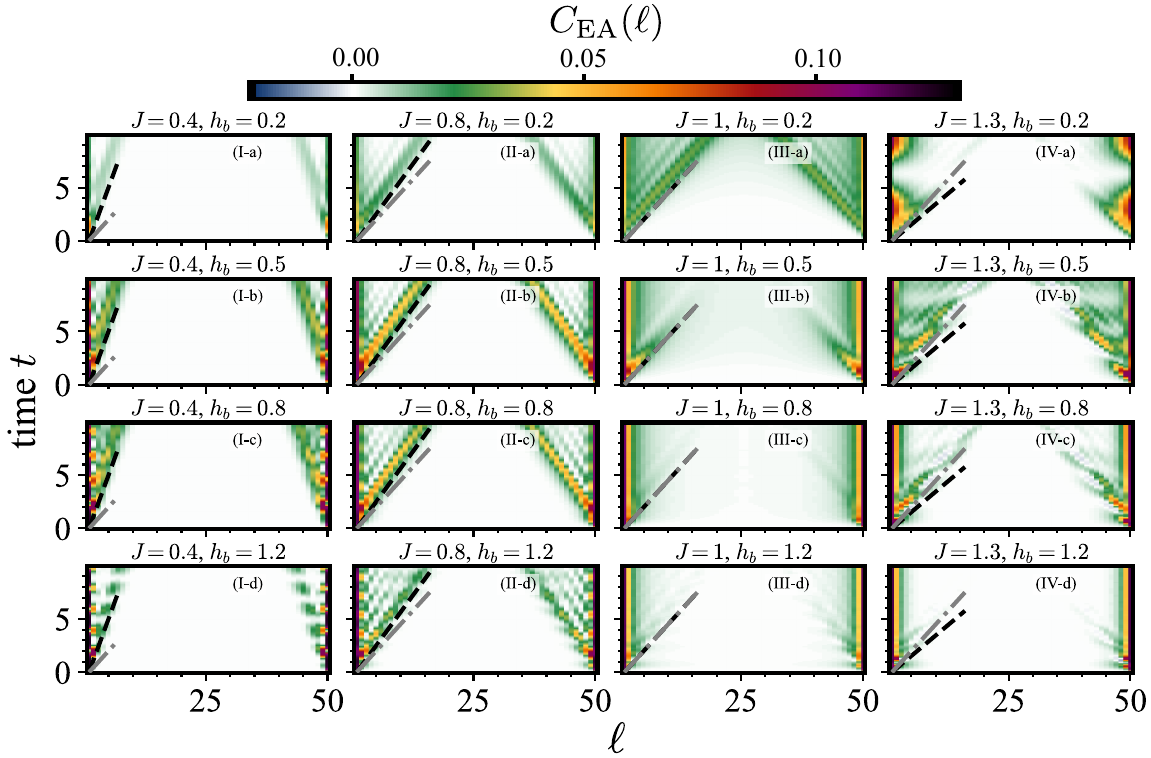}  
\end{center} 
\caption{
Color maps of $C_{\rm EA}(\ell)$ as a function of the site index $\ell$ and time $t$.
Each row corresponds to a different value of $h_b$ while each column corresponds to a different value of $J$.
The values of $h_b$ increase from top to bottom, whereas $J$ increases from left to right.
We set $L=50$.
The black dashed and gray dash-dotted lines indicate the propagation fronts corresponding to the quasiparticle velocity $2J$ and $2h$, respectively.
}
\label{Fig_6}
\end{figure*}
We have so far observed that $\Delta S_2$ is strongly influenced by the behavior of the boundary magnetization. More generally, because $\Delta S_2$ characterizes the overall symmetry-breaking information within a subsystem, it does not directly resolve how this information is distributed in space.

To obtain a spatially resolved characterization, we consider the difference in $\Delta S_2$ between two subsystems whose sizes differ by one site:
\begin{eqnarray}
I_2^A(\ell)=\Delta S_2(A_\ell)-\Delta S_2(A_{\ell-1})\nonumber
\end{eqnarray}
and
\begin{eqnarray}
I_2^B(\ell)=\Delta S_2(B_\ell)-\Delta S_2(B_{\ell+1})\nonumber
\end{eqnarray}
where $A_\ell=\{0,1,\cdots,\ell\}$ and $B_\ell=\{\ell,\ell+1,\cdots,L+1\}$ denote the subsystems extending from the left and right boundaries to the site $\ell$ [see Appendix~\ref{Append_F} for the calculation of $\Delta S_2(B_\ell)$].
Note that $I_2^{A}(1)=\Delta S_2(\{1\})$ and $I_2^{B}(L)=\Delta S_2(\{L\})$.
Since the two subsystems share most of their degrees of freedom, this difference captures the incremental change associated with adding one site while reducing contributions common to both, including the strong boundary contribution.
Inspired by the entanglement contour~\cite{Chen2014,Wen2020,MacCormack2023,Kuno2023}, we introduce the contour-like finite-difference quantity named the (R\'enyi-2) asymmetry contour: 
\begin{eqnarray}
C_{\rm EA}(\ell)=\frac{1}{2}(I_2^{A}(\ell)+I_2^{B}(\ell)).
\label{def_AC}
\end{eqnarray}
Notably, $C_{\rm EA}(\ell)$ is not guaranteed to be non-negative.
Nevertheless, $C_{EA}(\ell)$ serves as a computationally tractable measure to characterize the incremental change in symmetry-breaking information as the subsystem is enlarged by one site.

Figure~\ref{Fig_6} shows the time evolution of $C_{\rm EA}(\ell)$ after the boundary quench for several values of $J/h$.
For $J/h\le1$, the propagation front of $C_{\rm EA}(\ell)$ is generally consistent with the velocity $v_{\rm max}=2\min(J,h)$.
For large $h_b$, $C_{\rm EA}(\ell)$ exhibits oscillations that appear to follow those of the magnetization.
For $J/h>1$, while $C_{\rm EA}(\ell)$ with small $h_b$ exhibits characteristic oscillatory behavior at the boundary, the propagation front of $C_{\rm EA}(\ell)$ is instead described by the velocity $2h$.
Notably, for several parameter values [see, e.g., Fig.~\ref{Fig_6}(III-b) and (IV-d)], while its magnitude is relatively small, $C_{\rm EA}(\ell)$ becomes finite even before the arrival of the quasiparticle front, which may reflect the nonlocal nature of $C_{\rm EA}(\ell)$.
These results show that $C_{\rm EA}(\ell)$ provides a spatially resolved view of the symmetry-breaking information captured by $\Delta S_2$.
Its dynamics exhibits signatures of the underlying quasiparticle propagation, together with additional spatial and temporal structures that depend on the boundary field strength.

\section{Conclusion}
In this work, we investigated the ground-state and nonequilibrium behavior of the transverse-field Ising chain in the presence of $Z_2$ symmetry-breaking boundary magnetic fields.
We characterized symmetry breaking using both the $x$-component magnetization and the Rényi-2 EA, $\Delta S_2$.
To enable calculation for larger system sizes, we extended the ancilla-spin approach to efficiently evaluate $\Delta S_2$ in parity-broken states.

For the ground state, the magnetization exhibits distinct behaviors reflecting the ordered and disordered phases of the TFIM.
$\Delta S_2$ also shows phase-dependent behavior consistent with the bounds derived in this work, which constrain $\Delta S_2$ in terms of the sum of the squared correlators.
Since the spatial behavior of these correlators strongly depends on whether the system is in the ordered or disordered regime, $\Delta S_2$ exhibits distinct behavior in the two phases of the TFIM.
At criticality, the numerical scaling of the magnetization and $\mathcal Z_2$ is consistent with the corresponding field-theoretical predictions.

Turning to the boundary-quench dynamics, we found that the magnetization propagates from the boundaries into the bulk with a front velocity consistent with the maximum quasiparticle velocity.
In contrast, $\Delta S_2$ reflects the symmetry-breaking information encoded in the entire subsystem and therefore does not directly resolve its spatial distribution.
To obtain a spatially resolved characterization of this information, we introduced the asymmetry contour $C_{EA}(\ell)$, a contour-like finite-difference quantity.
The dynamics of $C_{\rm EA}(\ell)$ exhibits propagation structures consistent with the quasiparticle dynamics, while retaining sensitivity to the nonlocal correlations encoded in EA beyond the local magnetization.

An interesting direction for future work is to develop a quasiparticle picture for the boundary quench~\cite{Essler2016}.
Although the post-quench Hamiltonian is quadratic, the parity-broken state is non-Gaussian.
Therefore, a conventional quasiparticle picture may require some modification.
Recent field-theoretical studies of EA~\cite{Kusuki2025,Fossati2025} and local quenches~\cite{Bonsignori2023,Barad2025,fossati2026} may provide useful insights into such an extension.
Another natural extension is to consider opposite-sign boundary magnetic fields.
Since such boundary magnetic fields can induce
a magnet-to-kink crossover~\cite{Campostrini2015}, it would be interesting to investigate how this crossover is reflected in $\Delta S_2$ and the $C_{\rm EA}$.
Finally, it would be interesting to explore whether the bounds on $\Delta S_2$ derived in this work can be extended to broader classes of systems.
If non-linear quantities such as EA can generally be constrained by simpler correlation-function-based quantities, such relations could provide a useful route toward a deeper theoretical and experimental understanding of symmetry breaking from an information-theoretic perspective.

\section*{Acknowledgements}
This work is supported by JSPS KAKENHI: 
JP26K17056 (T.O.), JP23K25830 (DY), JP24K06890 (DY), JP26K00664 (DY), and JST PRESTO Grant No.~JPMJPR245D (DY).

\section*{Data availability}
The numerical code supporting the findings of this article is openly available on Zenodo~\cite{Orito2026Code}.

\appendix

\setcounter{section}{0}

\renewcommand{\thesection}{\Alph{section}}
\renewcommand{\theequation}{\thesection\arabic{equation}}
\renewcommand{\thefigure}{\thesection\arabic{figure}}
\renewcommand{\thetable}{\thesection\arabic{table}}

\makeatletter
\@addtoreset{equation}{section}
\@addtoreset{figure}{section}
\@addtoreset{table}{section}

\renewcommand{\theHsection}{\Alph{section}}
\renewcommand{\theHequation}{\thesection.\arabic{equation}}
\renewcommand{\theHfigure}{\thesection.\arabic{figure}}
\renewcommand{\theHtable}{\thesection.\arabic{table}}
\makeatother

\section{Calculation of $\langle G_+|X_{L+1}|G_+\rangle$}
\label{Append_A}
The expectation value of $X_{L+1}$ can be calculated using Wick's theorem.
Using Eq.~(\ref{X_jcalc}),
$\langle G_+| X_{L+1}|G_+\rangle$ can be reduced to $\langle G_+| X_0X_{L+1}|G_+\rangle=\langle {0}| X_0X_{L+1}|{0}\rangle$, where $|{0}\rangle$ is the Bogoliubov vacuum.
To express $X_0X_{L+1}$ in a compact form, we introduce
\begin{eqnarray}
C_i=c_i^\dagger+c_i,\ D_i=c_i^\dagger-c_i.
\end{eqnarray}
Using these two operators, $X_0X_{L+1}$ can be written as 
\begin{eqnarray}
X_0X_{L+1}=D_0C_1D_1\cdots C_{L-1}D_LC_{L+1}.
\end{eqnarray}
The calculation of Wick’s theorem can be reduced to the calculation of Pfaffian: $\langle X_0X_{L+1}\rangle={\rm Pf}[E^{(L+1)}]$, where
\begin{equation}
E^{(i)}_{mn}
=
\begin{cases}
\left\langle O^{(i)}_m O^{(i)}_n \right\rangle,
& m<n,\\
-\left\langle O^{(i)}_n O^{(i)}_m \right\rangle,
& m>n,\\
0,
& m=n.
\end{cases}
\label{Pfaffian}
\end{equation}
Here,
\begin{equation}
\bm{O}^{(i)}
=
\left(
D_0,\,
C_1,\,
D_1,\,
\ldots,\,
C_{i-1},\,
D_{i-1},\,
C_i
\right)^{T}.
\end{equation}

The expectation value of $X_i$ can be calculated in a similar manner.

\section{Perturbative analysis for paramagnetic and ferromagnetic phases}
\label{Append_B}

\begin{figure}[t]
\begin{center} 
\vspace{0.5cm}
\includegraphics[width=7cm]{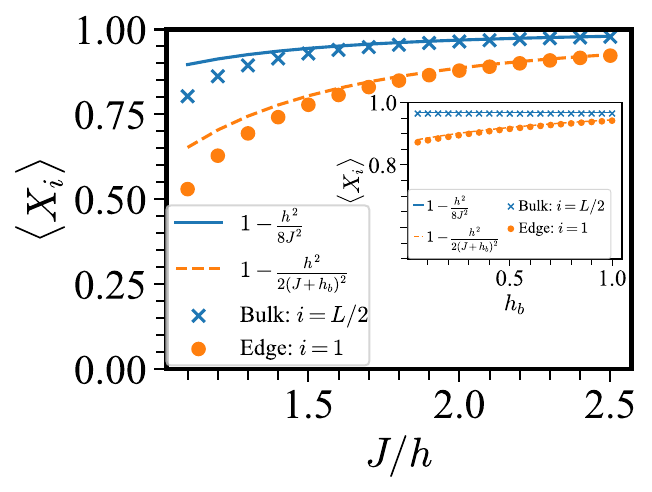}  
\end{center} 
\caption{Numerical result for bulk ($i=L/2$) and edge ($i=1$) magnetization as a function of $J/h$.
We set $h_b=0.1$, $h=1$, and $L=100$.
The inset illustrates bulk ($i=L/2$) and edge ($i=1$) magnetization as a function of $h_b$. We set $J=2$, $h=1$, and $L=100$.
}
\label{AppendFig_1}
\end{figure}

In a paramagnetic phase $J<<h$, we regard the Ising interaction and boundary magnetic fields as perturbation terms.
First, assuming $J=0$ and $h_b=0$, a unperturbative ground state can be written as 
$|\psi^{0}\rangle=\prod_i|\uparrow_i\rangle$, where $Z_i|\uparrow_i\rangle=|\uparrow_i\rangle$.
It is easy to show that $\langle\psi^0|X_i|\psi^0\rangle=0\ {\rm for\ all \ } X_i$.
A finite value of $\langle X_i\rangle$ requires a perturbative path connecting site $i$ to either the local symmetry-breaking boundary magnetic fields.
This is because bulk perturbation terms $X_iX_{i+1}$ preserve $Z_2$ symmetry, which indicates that it alone cannot lead to $\langle X_i\rangle\neq0$.
Taking into account both boundary's perturbations, the leading contribution appears at order ${h_{b}\over h}({J\over h})^{i-1}+{h_{b}\over h}({J\over h})^{L-i}$, yielding $m_i\propto {h_{b}\over h}({J\over h})^{i-1}+{h_{b}\over h}({J\over h})^{L-i}={h_{b}\over h} e^{-{(i-1)\over\xi}}+{h_{b}\over h}e^{-{(L-i)\over\xi}}$, where $\xi=1/\ln(h/J)$.
$\xi=1/\ln(h/J)$ obtained from the perturbative analysis is consistent with the result presented in the main text.

In a ferromagnetic phase $J>>h$, we regard the transverse magnetic fields $H_g$ as the perturbation term, and we denote the Ising coupling and boundary magnetic fields as $H_0$.
For $h=0$,  a unperturbative ground state can be written as 
$|\psi^{0}\rangle=|+\rangle=\prod_i|\rightarrow_i\rangle$, where $X_i|\rightarrow_i\rangle=|\rightarrow_i\rangle$.
The first-order (unnormalized) perturbative eigenstate is given by
\begin{eqnarray}
|\Psi\rangle&=&|+\rangle+|\chi\rangle\nonumber\\
&=&(1+\alpha[Z_1+ Z_L]+\beta\sum_{i=2}^{L-1}Z_i)|+\rangle+\mathcal{O}(h^2).\nonumber\\
\label{perturvation1}
\end{eqnarray}
Here, the coefficients $\alpha$ and $\beta$ are determined by the following eigenvalue condition,
\begin{eqnarray}
H_g|+\rangle=(E_0-H_0)(|\Psi\rangle-|+\rangle),
\label{perturvation2}
\end{eqnarray}
where $E_0=-J(L-1)-2h_b$ is the eigenenergy of the unperturbative state, and we here neglect $\mathcal{O}(h^2)$ contributions to the eigenstate and energy
~\footnote{The eigenvalue problem can be written as 
$(H_0+H_g)(|+\rangle+|\chi\rangle)=E_0(|+\rangle+|\chi\rangle)+\mathcal{O}(h^2)$.
Neglecting $\mathcal{O}(h^2)$ contributions, we obtain
$H_g|+\rangle=(E_0-H_0)|\chi\rangle=(E_0-H_0)(|\Psi\rangle-|+\rangle)$.}.
By comparing with the left-hand and the right-hand sides of Eq.~(\ref{perturvation2}),
we obtain 
\begin{eqnarray}
\alpha={h\over2(J+h_b)} {\ \rm and}\ \beta={h\over4J}.
\label{perturvation3}
\end{eqnarray}
Using Eq.~(\ref{perturvation3}), $\langle X_i\rangle$ can be calculated as 
\begin{widetext}
\begin{eqnarray}
\langle X_{i\neq1,L}\rangle&=&{\langle\Psi|X_i|\Psi\rangle\over\langle\Psi|\Psi\rangle}={1+{h^2\over2(J+h_b)^2}+{h^2(L-4)\over16J^2}\over1+{h^2\over2(J+h_b)^2}+{h^2(L-2)\over16J^2}}\sim 1-{h^2\over8J^2}+\mathcal{O}(h/J)^4.
\label{Xi_for_ineq1,L}
\end{eqnarray}
and
\begin{eqnarray}
\langle X_{i=1,L}\rangle&=&{\langle\Psi|X_i|\Psi\rangle\over\langle\Psi|\Psi\rangle}={1+{h^2(L-2)\over16J^2}\over1+{h^2\over2(J+h_b)^2}+{h^2(L-2)\over16J^2}}\sim 1-{h^2\over2(J+h_b)^2}+\mathcal{O}(h/J)^4.
\label{Xi_for_1,L}
\end{eqnarray}
\end{widetext}

Figure~\ref{AppendFig_1} shows the $x$-component of bulk and edge magnetization as a function of $J/h$.
As $J/h$ increases, numerical results show better agreement with the perturbative analysis.
In the inset of Fig.~\ref{AppendFig_1}, we also confirm that the $h_b$ dependence of $\langle X_i\rangle$ is consistent with the perturbative analysis as well.
Notably, the bulk magnetization Eq.~(\ref{Xi_for_ineq1,L}) is close to the leading order behavior obtained from the exact spontaneous magnetization of the TFIM $\langle X_i\rangle\sim(1-(h/J)^2)^{1/8}\sim 1-{h^2\over8J^2}+\mathcal{O}(h/J)^4$~\cite{PFEUTY197079}. 

\section{Derivation of Eq.~(\ref{commutation_rho_0})}
\label{Append_C}
To derive $[\rho_A^0, X_0]=0$, it is instructive to work in the Majorana basis.
Following the well-established Peschel formula\cite{Peschel2009,Fagotti2010,Bravyi2005}, $\rho_A^0$ can be written as
\begin{eqnarray}
\rho_A^0=\prod_{n=0}^{\ell_A}{1\over2}[1+i\nu_n\tilde{a}_{2n}\tilde{a}_{2n+1}],
\end{eqnarray}
where $i\nu_n$, $\tilde{a}_{2n}$ and $\tilde{a}_{2n+1}$ are eigenvalues and canonical Majorana modes of $\Gamma$, respectively.
Canonical Majorana modes can be obtained by bringing the Majorana covariance matrix $\Gamma$ into its canonical form and are written as 
\begin{eqnarray}
\begin{pmatrix}
   \tilde{a}_0 \\ 
   \vdots \\
   \tilde{a}_{\ell_A}\\
   \tilde{a}_{\ell_A+1} \\ 
   \vdots \\
   \tilde{a}_{2\ell_A+1}
\end{pmatrix}
=R
\begin{pmatrix}
   {a}_0 \\ 
   \vdots \\
   {a}_{\ell_A}\\
   {a}_{\ell_A+1} \\ 
   \vdots \\
   {a}_{2\ell_A+1}
\end{pmatrix},
\label{canonical transformation}
\end{eqnarray}
where $R$ is the real orthogonal matrix.

Let us focus on the property of $\rho_A^0$.
Here, it is instructive to examine the first row of the correlation matrix, i.e., $\Gamma_{0,j}$.
One notices that $\Gamma_{0,j}=0$ for $j\neq0$ because
\begin{widetext}
\begin{eqnarray}
\langle a_0 a_{j}\rangle&=&\langle (\eta_0+\eta_0^\dagger)(\sum_{n=1}^{L+1}f_{n,j}^*\eta_n+f_{n,j}\eta_n^\dagger)\rangle
=\langle \eta_0(\sum_{n=1}^{L+1}f_{n,j}^*\eta_n+f_{n,j}\eta_n^\dagger)\rangle
=\langle -(\sum_{n=1}^{L+1}f_{n,j}^*\eta_n+f_{n,j}\eta_n^\dagger)\eta_0\rangle=0,
\label{a_decouple}
\end{eqnarray}
where the coefficients $f_{n,j}$ are determined from Eq.~(\ref{annihilation and creation}).
Equation~(\ref{a_decouple}) indicates that $a_0$ ($=X_0$) is the canonical eigenmode $\tilde{a}_0$ whose eigenvalue is zero, and $R_{0,j}=\delta_{0,j}$.
It is now straightforward to show that 
\begin{eqnarray}
\{X_0,\tilde{a}_{i\neq0}\}&=&\{a_0,\sum_j R_{{i\neq0},j} a_j\}
=\sum_j R_{{i\neq0},j}\{a_0,a_j\}
=2R_{{i\neq0},0}
=0,
\end{eqnarray}
and for $n\neq0$, we obtain
\begin{eqnarray}
[X_0,\tilde{a}_{2n}\tilde{a}_{2n+1}]&=&0.
\label{commutation relation of X_0}
\end{eqnarray}
Using Eq.~(\ref{commutation relation of X_0}) and the fact that $\nu_0=0$, we can show that 
\begin{eqnarray}
[X_0,\rho_A^0]&=&X_0\prod_{n=0}^{\ell_A}{1\over2}[1+i\nu_n\tilde{a}_{2n}\tilde{a}_{2n+1}]-\prod_{n=0}^{\ell_A}{1\over2}[1+i\nu_n\tilde{a}_{2n}\tilde{a}_{2n+1}]X_0\nonumber\\
&=&{X_0\over2}\prod_{n=1}^{\ell_A}{1\over2}[1+i\nu_n\tilde{a}_{2n}\tilde{a}_{2n+1}]-\prod_{n=1}^{\ell_A}{1\over2}[1+i\nu_n\tilde{a}_{2n}\tilde{a}_{2n+1}]{X_0\over2}=0.
\end{eqnarray}
\end{widetext}

\section{Derivation of Eq.~(\ref{def_REA3})}
\label{Append_D}
To derive Eq.~(\ref{def_REA3}), we first rewrite $I-\Gamma\Gamma'$ in a form suitable for applying the matrix determinant lemma: 
\begin{eqnarray}
I-\Gamma\Gamma'&=&I-\Gamma R\Gamma R^T\nonumber\\
&=&G+XY^T,
\label{Mat-det1}
\end{eqnarray}
where $G=I-\Gamma^2$, $X=\begin{pmatrix}-2\Gamma e_{\ell_A+1}  &2\Gamma^2 e_{\ell_A+1}\end{pmatrix}$, and  $Y=\begin{pmatrix}\Gamma e_{\ell_A+1}  &e_{\ell_A+1}\end{pmatrix}$.
To derive Eq.~(\ref{Mat-det1}), we use the fact $\Gamma$ is an anti-symmetric matrix, i.e., $e_{\ell_A+1}^T \Gamma e_{\ell_A+1}=0$. 
Using the matrix determinant lemma~\cite{Harville1997}:
\begin{eqnarray}
\det(G+XY^T)=\det(G)\det(I_2+Y^TG^{-1}X),
\label{Mat-det2}
\end{eqnarray}
we obtain 
\begin{eqnarray}
\sqrt{\det[I-\Gamma \Gamma']\over\det[I-\Gamma^2]}=\sqrt{\det(I_2+Y^TG^{-1}X)}.
\label{Mat-det3}
\end{eqnarray}
Equation~(\ref{Mat-det3}) can be further simplified:
\begin{eqnarray}
Y^TG^{-1}X
&=&
\begin{pmatrix}
2q-2& 0\\
0& 2q-2
\end{pmatrix},
\label{Mat-det4}
\end{eqnarray}
where $q=e_{\ell_A+1}^TG^{-1} e_{\ell_A+1}=[(I-\Gamma^2)^{-1}]_{\ell_{A+1},\ell_{A+1}}$.
Note that $q\ge1/2$ since the eigenvalues of $-\Gamma^2$, $\lambda_i$ satisfy $0\le\lambda_i\le1$.
To derive Eq.~(\ref{Mat-det4}), we use the fact that $G^{-1}\Gamma^3$ and $G^{-1}\Gamma$ are anti-symmetric matrix, and $G^{-1}\Gamma^2=G^{-1}-I$. 
Using Eq.~(\ref{Mat-det4}), we now arrive
\begin{eqnarray}
\Delta S_2&=&-\ln[{1\over2}(1+\sqrt{\det(I_2+Y^TG^{-1}X)})]\nonumber\\
&=&\ln2-\ln[(1+(2q-1))]\nonumber\\
&=&-\ln[(I-\Gamma^2)^{-1}_{\ell_A+1,\ell_A+1}].
\end{eqnarray}

\section{Derivation of Eq.~(\ref{def_inequality1}) and Eq.~(\ref{def_inequality2})} 
\label{Append_E}
To derive Eq.~(\ref{def_inequality1}), first, note that $F\equiv-\Gamma^2$ is a Hermitian positive semidefinite matrix and admits the spectral decomposition as 
\begin{eqnarray}
F=O\ {\rm diag \lambda_i}\ O^\dagger, 
\end{eqnarray}
where $O$ is a unitary matrix and $0\le\lambda_i\le1$.
It is easy to show that 
\begin{eqnarray}
(1+F)^{-1}_{\ell_A+1,\ell_A+1}=\sum_i {|O_{\ell_A+1,i}|^2\over 1+\lambda_i}.
\end{eqnarray}
\begin{widetext}
We then define
\begin{eqnarray}
s&\equiv&F_{\ell_A+1,\ell_A+1}=\sum_i|\Gamma_{\ell_A+1,i}|^2=\sum_i|O_{\ell_A+1,i}|^2\lambda_i
\end{eqnarray}
The key observation is that $f(\lambda)={1\over1+\lambda}$ is convex for $\lambda\ge0$, since $f''(\lambda)\ge0$ and $\sum_i|O_{\ell_A+1,i}|^2=1$.
Therefore, using Jensen's inequality, we obtain
\begin{eqnarray}
(1+F)^{-1}_{\ell_A+1,\ell_A+1}=\sum_i {|O_{\ell_A+1,i}|^2}f(\lambda_i)\ge f(\sum_i|O_{\ell_A+1,i}|^2\lambda_i)={1\over1+s}.
\label{Jensen_ineq}
\end{eqnarray}
\end{widetext}
Using Eq.~(\ref{Jensen_ineq}), it is now straightforward to show that
\begin{eqnarray}
(1+F)^{-1}_{\ell_A+1,\ell_A+1}\ge{1\over1+s}
\Rightarrow \Delta S_2\le\ln(1+s).\nonumber\\
\label{upperbound}
\end{eqnarray}
Regarding a lower bound of $\Delta S_2$, we use the following inequality:
\begin{eqnarray}
{1\over1+\lambda}\le1-{\lambda\over2}, 
\end{eqnarray}
where $0\le\lambda\le1$.
Therefore, we can show that
\begin{eqnarray}
\sum_i {|O_{\ell_A+1,i}|^2\over 1+\lambda_i}\le \sum_i |O_{\ell_A+1,i}|^2(1-{\lambda_i\over2})=1-{s\over2}.
\label{lowerbound}
\end{eqnarray}
Using Eq.~(\ref{lowerbound}), we now arrive a lower bound of $\Delta S_2$:
\begin{eqnarray}
\sum_i {|O_{\ell_A+1,i}|^2\over 1+\lambda_i}\le 1-{s\over2}\Rightarrow \Delta S_2\ge -\ln(1-{s\over2}).
\end{eqnarray}

We now turn to our focus on the derivation of Eq.~(\ref{def_inequality2}).
To this end, it is useful to rewrite the squared expectation value as
\begin{eqnarray}
\langle X_0X_i\rangle^2={\rm Pf}[E^{(i)}] ^2=\det[E^{(i)}]\ge0. 
\end{eqnarray}
Here, we use the well-known relationship between the Pfaffian and the determinant of a skew matrix~\cite{Wimmer2012}.
The singular value of $E^{(i)}$ appears in pairs $(\mu_j,\mu_j)$,  $0\le\mu_j\le1$, and therefore 
\begin{eqnarray}
\det[E^{(i)}]=\prod_{j=0}^{i-1} \mu_j^2.
\end{eqnarray}
Let $\mu_0$ denote the smallest singular value.
We then obtain the following inequality: 
\begin{eqnarray}
\det[E^{(i)}]=\prod_{j=0}^{i-1} \mu_j^2\le\mu_{0}^2.
\end{eqnarray}
Here, as a direct consequence of the definition of $\mu_0$, we have 
\begin{eqnarray}
\mu_0=\underset{||v||=1}{\rm min}||E^{(i)}v||\le ||E^{(i)}e_1||,
\end{eqnarray}
where $e_1$ denotes the unit vector in the $1$ direction.
Then, we obtain 
\begin{eqnarray}
\langle X_0X_i\rangle^2\le ||E^{(i)}e_1||^2=\sum_{j=1}^{i-1}|\langle D_0D_j\rangle|^2+\sum_{j=1}^{i}|\langle D_0C_j\rangle|^2.\nonumber\\
\end{eqnarray}
Furthermore, since $s$ can be written as  
\begin{eqnarray}
s=\sum_i|\Gamma_{\ell_A+1,i}|^2=\sum_{j=1}^{\ell_A}|\langle D_0D_j\rangle|^2+\sum_{j=0}^{\ell_A}|\langle D_0C_j\rangle|^2,\nonumber\\
\end{eqnarray}
it is straightforward to show that 
\begin{eqnarray}
\langle X_0X_i\rangle^2\le ||E^{(i)}e_1||^2\le s.
\label{inequ_Xvss}
\end{eqnarray}
By substituting Eq.~(\ref{inequ_Xvss}) into Eq.~(\ref{def_inequality1}), we arrive
\begin{eqnarray}
-\ln(1-{1\over2}\underset{1\le i\le\ell_A}{\rm max}\langle X_0X_i\rangle^2) \le -\ln(1-{1\over2}s)\le\Delta S_2.\nonumber\\
\end{eqnarray}

\section{Calculation of $\Delta S_2(B_\ell)$}
\label{Append_F}
\begin{figure}[t]
\begin{center} 
\vspace{0.5cm}
\includegraphics[width=7cm]{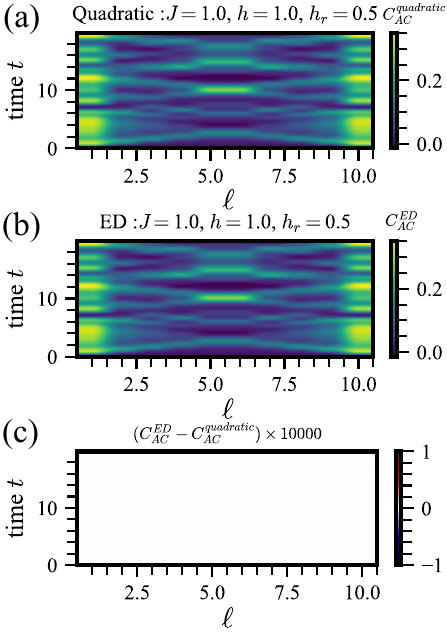}  
\end{center} 
\caption{Color maps of $C_{\rm EA}(\ell)$ as a function of the site index $\ell$ and time $t$.
(a) Numerical result obtained from the quadratic-fermion calculation.
(b) Result obtained using exact diagonalization. 
(c) Difference between the results in panels (a) and (b). 
The system size is fixed at $L=10$. $J/h=1$ and $h_b=0.5$.
}
\label{AppendFig_2}
\end{figure}

We briefly describe the fermionic calculation of $\Delta S_2(B_\ell)$.
To this end, we first calculate the reduced density matrix for subsystem $B$.
Following Refs.~\cite{Abdul2015,Fagotti2010}, we introduce fermions local to subsystem $B$,
\begin{eqnarray}
c_i^{b}=(\prod_{j=\ell}^{i-1}Z_j)\sigma^-_i,
\end{eqnarray}
where $\sigma^-_i=(X_i-iY_i)/2$.
They are related to the original fermions through
\begin{eqnarray}
c_i=P_Ac_i^b,
\end{eqnarray}
where $P_A=\prod_{i=0}^{\ell-1}Z_i$ and is related to the Parity operator $P$ : $P=P_AP_B$ with  $P_B=\prod_{i=\ell}^{L+1}Z_i$.
We choose the Bogoliubov vacuum state $|0\rangle$ in the even-parity sector, such that
\begin{eqnarray}
P_AP_B|0\rangle=|0\rangle,
\end{eqnarray}
and consider 
\begin{eqnarray}
|G_+\rangle&=&{1\over\sqrt{2}}(1+\eta^\dagger_0)|0\rangle\nonumber\\
&=&{1\over\sqrt{2}}(1+\eta^\dagger_0-\eta_0)|0\rangle\nonumber\\
&=&{1\over\sqrt{2}}(1+P_Aq_R)|0\rangle\nonumber\\
&=&{1\over\sqrt{2}}(1+q_RP_B)|0\rangle,
\end{eqnarray}
where $q_R=c^{b\dagger}_{L+1}-c^{b}_{L+1}$ and we use $P_A|0\rangle=P_B|0\rangle$ in the fourth line.
The reduced density matrix for the right subsystem can be written as 
\begin{eqnarray}
\rho_B={1\over2}(1+q_RP_B)\rho_B^0(1+q_RP_B),
\end{eqnarray}
where $\rho_B^0=\Tr_A[|0\rangle\langle0|]$.

Similarly to $\rho_A$, $\rho_B^0$ commutes with the Parity-odd operator $q_RP_B$. 
To see this, first, note that $X_{L+1}=\prod_{i=0}^{L}(1-2n_i)(c_{L+1}^\dagger+c_{L+1})=q_RP_B$ and $X_0X_{L+1}|0\rangle=|0\rangle$.
Using these two identities, it is straightforward to show that 
\begin{eqnarray}
q_RP_B\rho_B^0q_RP_B&=&\Tr_A[q_RP_B|0\rangle\langle0|q_RP_B]\nonumber\\
&=&\Tr_A[X_0|0\rangle\langle0|X_0]\nonumber\\
&=&\rho^0_B.
\end{eqnarray}
Consequently, $\rho_B$ can be simplified: 
\begin{eqnarray}
\rho_B=(1+q_RP_B)\rho_B^0.
\end{eqnarray}
As in the calculation for the subsystem $A$,
we introduce the Majorana operators for the subsystem $B$:
\begin{eqnarray}
\Phi^B=
\begin{pmatrix}
   {a}_{0}^B \\ 
   \vdots \\
   {a}_{L+1-\ell}^B\\
   {a}_{L+2-\ell}^B \\ 
   \vdots \\
   {a}_{2L+3-2\ell}^B
\end{pmatrix}
=
\begin{pmatrix}
   c_\ell^b+c_\ell^{b\dagger} \\ 
   \vdots \\
   c_{L+1}^b+c_{L+1}^{b\dagger}\\
   i(c_\ell^{b\dagger}-c_\ell^b) \\ 
   \vdots \\
   i(c_{L+1}^{b\dagger}-c_{L+1}^b)\\
\end{pmatrix}.
\end{eqnarray}
We then define the restricted covariance matrix $M^B$ for the subsystem $B$:
\begin{eqnarray}
M_{i,i'}^B=\langle {a}^B_i {a}^B_{i'} \rangle=\delta_{i,i'}+i\Gamma_{i,i'}^B,
\end{eqnarray}
where $(\Gamma^B)^{T}=-\Gamma^B$.
As with $\rho^0_A$, $\rho^0_B$ is completely determined by $\Gamma^B$.
We note that, since $c_i^b=P_A c_i$ and $[P_A,c_i^b]=[P_A,c_i^{b\dagger}]=0$, the bilinear products coincide:
\begin{eqnarray}
c_i^b c_j^b&=&c_i c_j\\
c_i^{b\dagger} c_j^b&=&c_i^\dagger c_j.
\end{eqnarray}
Consequently, the two-point correlation functions of the fermions local to the subsystem $B$ are identical to those of the original fermions.

Using the Gaussian overlap formula and the matrix determinant lemma, we can show that 
\begin{eqnarray}
\Delta S_2(B_\ell)=-\ln([I-(\Gamma^B)^2]^{-1}_{L-\ell+1,L-\ell+1}).
\label{EA_B}
\end{eqnarray}
Since the parity-broken structure is preserved under time evolution, as in the case of $\rho_A$, the above construction of $\rho_B$ applies to the quench dynamics and can be used to evaluate $\Delta S_2(B_\ell)$.

To verify the validity of Eqs.~(\ref{def_REA3}) and (\ref{EA_B}), we compare the numerical results for the asymmetry contour $C_{\rm EA}(\ell)$ (\ref{def_AC}) with those obtained using the exact diagonalization (ED) method implemented with the QuSpin library~\cite{QuSpin1,Quspin2}.
Figure~\ref{AppendFig_2} shows the time evolution of $C_{\rm EA}(\ell)$.
The results obtained from quadratic-fermion calculation (panel (a)) show good agreement with the ED results (panel (b)).
As shown in panel (c), the difference between the two results is negligibly small within the numerical accuracy, confirming the validity of Eqs.~(\ref{def_REA3}) and (\ref{EA_B}).
\clearpage

\bibliography{ref}
\end{document}